# Robust Differential Received Signal Strength-Based Localization

Yongchang Hu, *Member, IEEE*, and Geert Leus, *Fellow, IEEE*

*Abstract*—Source localization based on signal strength measurements has become very popular due to its practical simplicity. However, the severe nonlinearity and non-convexity make the related optimization problem mathematically difficult to solve, especially when the transmit power or the path-loss exponent (PLE) is unknown. Moreover, even if the PLE is known but not perfectly estimated or the anchor location information is not accurate, the constructed data model will become uncertain, making the problem again hard to solve. This paper particularly focuses on differential received signal strength (DRSS)-based localization with model uncertainties in case of unknown transmit power and PLE. A new whitened model for DRSS-based localization with unknown transmit powers is first presented and investigated. When assuming the PLE is known, we introduce two estimators based on an exact data model, an advanced best linear unbiased estimator (A-BLUE) and a Lagrangian estimator (LE), and then we present a robust semidefinite programming (SDP)-based estimator (RSDPE), which can cope with model uncertainties (imperfect PLE and inaccurate anchor location information). The three proposed estimators have their own advantages from different perspectives: the A-BLUE has the lowest complexity; the LE holds the best accuracy for a small measurement noise; and the RSDPE yields the best performance under a large measurement noise and possesses a very good robustness against model uncertainties. Finally, we propose a robust SDP-based block coordinate descent estimator (RSDP-BCDE) to deal with a completely unknown PLE and its performance converges to that of the RSDPE using a perfectly known PLE.

*Index Terms*—Source localization, differential received signal strength (DRSS), path-loss exponent (PLE), least squares, Lagrangian multiplier, semidefinite programming (SDP), convex optimization, block coordinate descent.

## I. INTRODUCTION

PRESENTLY source localization is a rather prevalent technique aimed at locating a target based on measurements related to pre-deployed distributed sensors with *prior* known locations [1], i.e., anchor nodes. Briefly speaking, the commonly used measurements include, for example, time-of-arrival (TOA), time-difference-of-arrival (TDOA), angle-of-arrival (AOA) and signal strength. Among those, signal strength, such as received signal strength (RSS) [2] and differential RSS (DRSS) [3], gradually becomes the primary concern of numerous engineers owing to its implementation simplicity. Compared with other kinds of measurements, employing the signal strength as a measurement requires neither clock synchronization as for TOA-based or TDOA-based localization nor an antenna array which is indispensable for AOA-based localization. Therefore, this kind of source localization is more cost-effective in terms of both hardware and software. Besides, sensors usually have very scarce resources like limited computational abilities, constrained communication capabilities and depletable batteries, which further emphasizes its significance.

The signal strength measurement is determined by the signal power after successful demodulation [4], [5], which is still subject to a complicated radio propagation channel [6]. Without elaborating on the details, observe that the term "RSS", in most literature, actually refers to the *large-scale* fading, the average of the instantaneous received signal power over several consecutive time slots, such that the *small-scale* fading, which is usually considered to be *Rayleigh* [7] or *Nakagami* [8] distributed, can be neglected. Please also refer to Appendix A for details on the RSS collection. Based on such an underlying assumption, the *log-normal* shadowing model can be used to characterize the RSS. Therefore, in $\mathbb{R}^d$, the RSS between the $i$-th anchor node, located at $\mathbf{s}_i$, and the target node, located at $\mathbf{x}$, can be presented in dB by

$$P_i = P_{0,i} - 10\gamma log_{10}\left(\frac{||\mathbf{x} - \mathbf{s}_i||_2}{d_0}\right) + \chi_i, \ i = 1, 2, \cdots, N, \quad (1)$$

where $P_{0,i}$ is the received power related to the $i$-th anchor node at the reference distance $d_0$, $\gamma$ is the path-loss exponent (PLE), $\chi_i \sim \mathcal{N}(0, \sigma_\chi^2)$ represents the shadowing effect and $N$ is the number of anchor nodes. Without loss of generality, we assume $d_0 = 1$ m. Note that $P_{0,i}$ can also be considered to be equivalent to the transmit power of the RSS related to the $i$-th anchor node. Then, the ultimate goal is to estimate the target location $\mathbf{x}$ from the RSS samples $P_i$ and known anchor locations $\mathbf{s}_i$.

To achieve this goal, source localization techniques using RSS measurements can be divided into three categories: maximum likelihood (ML), least squares (LS) based and semidefinite programming (SDP) based. The ML method is asymptotically optimal, but the related ML optimization problem is highly non-linear and non-convex [9]. Admittedly speaking, it can be

Manuscript received January 13, 2016; revised August 1, 2016 and February 2, 2017; accepted March 2, 2017. Date of publication March 20, 2017; date of current version April 24, 2017. The associate editor coordinating the review of this manuscript and approving it for publication was Dr. Fauzia Ahmad. This work was supported by the China Scholarship Council and the Circuits and Systems group, Delft University of Technology, The Netherlands. *(Corresponding author: Yongchang Hu.)*

The authors are with the Department of Microelectronics, Delft University of Technology, Delft 2628CD, The Netherlands (e-mail: y.hu-1@tudelft.nl; g.j.t.leus@tudelft.nl).

Color versions of one or more of the figures in this paper are available online at http://ieeexplore.ieee.org.

Digital Object Identifier 10.1109/TSP.2017.2684741

iteratively solved [10]–[14]. However, this actually comes at the price of a high computational complexity. Moreover, the non-convexity also implies multiple local minima and hence an appropriate initialization is very important. The LS-based method relies on tackling the non-linearity by converting the non-linear ML optimization problem into a linear form such that some (weighted) LS-based solutions can be easily obtained [15]–[18]. However, these estimators are very susceptible to a large shadowing effect. The SDP-based method deals with the non-convexity by relaxing the non-convex optimization problem to a convex one such that a global minimum can be effectively found [15], [19]–[23]. However, this method still requires a high complexity as well as a tight relaxation to guarantee an accurate estimate.

Besides those aforementioned issues, it is worth noting that the RSS measurements can be collected either by anchors in a distributed fashion or locally by the target node. To be specific, the former indicates that the localization signal is broadcast only by the target node and hence the transmit power $P_{0,i}, \forall i$ is obviously the same for all the RSS measurements. However, in the latter case, when several localization signals are emitted by the anchors, the related transmit power $P_{0,i}, \forall i$ should be considered different. This is because, even if anchors are equipped with stable and sustainable power supplies to guarantee a consistent transmit power, a deviation $\Delta P_{0,i}$ can still occur due to some unexpected power surges or system instabilities and hence we have $P_{0,i} \triangleq \bar{P}_0 + \Delta P_{0,i}$ with $\bar{P}_0$ the nominal transmit power. Besides, some transmit power control techniques are often carried out for energy saving purpose, which could also result in a $\Delta P_{0,i}$. Compared with the former case, which might require particular networking protocols to aggregate the collected RSS measurements to a computation center (CC) for localization, the latter is more convenient and widely assumed, since the target node can just listen and then self-estimate its location based on the locally collected RSS measurements without increasing any workload related to the wireless networking. However, to the best of our knowledge, current RSS-based localization techniques rarely consider the case of different $P_{0,i}$.

In either one of the aforementioned cases, if the network is not very cooperative or the signal transmitter intentionally withholds information (e.g., for military scenarios), $P_{0,i}$ will be unknown. Similarly, the PLE $\gamma$ is often unknown as well, since it might be very difficult or expensive to acquire, especially in dynamic communication environments. Yet, many works simply assume that they are perfectly known [12], [19], [20], [23], [24]. To tackle the problem of an unknown PLE $\gamma$, a pre-calibration procedure of the PLE can be carried out among the anchor nodes before the actual localization phase [25]–[28]. However, this will consume extra resources and will make the implementation more cumbersome. Consequently, some joint estimators of $\mathbf{x}$ and $\gamma$ appear in [10], [17], [18], [22], [29]–[32]. To handle the issue of an unknown $P_{0,i}$, there are also some joint estimators for $\mathbf{x}$ and $P_{0,i}$ [15], [16], [21], [22], [30], [31].

In this paper, instead of utilizing RSS measurements, we consider DRSS measurements for localization. The practical advantages for using the DRSS measurements are similar to those of TDOA-based localization. While preserving all the advantages of RSS-based localization, it can significantly alleviate the passive dependence of localization on the signal transmitter, which could be defective, malicious or uncooperative. Moreover, control overhead message between anchors and target node is minimized or even no longer required, which saves energy, bandwidth and throughput. This also conceals the localization process from the signal transmitter, which is very beneficial to surveillance or military applications. Therefore, DRSS-based localization is very promising. Considering different unknown transmit powers $P_{0,i}$, the DRSS measurements can be obtained from (1) as

$$P_{i,1} = -10\gamma log_{10}\left(\frac{||\mathbf{x} - \mathbf{s}_i||_2}{||\mathbf{x} - \mathbf{s}_1||_2}\right) + \Delta P_{0,i,1} + \chi_{i,1}, \ i \neq 1, \quad (2)$$

where $P_{i,1} = P_i - P_1$, $\Delta P_{0,i,1} \triangleq \Delta P_{0,i} - \Delta P_{0,1}$ and $\chi_{i,1} = \chi_i - \chi_1$. To construct a DRSS sample set, a reference node (RN) is chosen and the measurements are taken w.r.t. that RN. For convenience, the RN is appointed as the first anchor. Note that, in such a case, the size of the DRSS sample set becomes $N - 1$. In spite of the fact that (2) still remains non-linear and non-convex, the benefit of using a DRSS sample set is that the unknown nominal transmit power $\bar{P}_0$ vanishes. However, compared with (1), the inevitable price is that the shadowing effect and the transmit power deviation are exacerbated since $\chi_{i,1}$ and $\Delta P_{0,i,1}$ become correlated and (2) gets even more complicated to solve. This is also the reason why very few papers study this type of localization. To the best of our knowledge, some early results occurred in [33], [34]. In [15], some initial DRSS-based localization techniques were presented, yet having a worse accuracy than the corresponding RSS-based localization techniques, except for a simple least squares (LS) estimator which merely is slightly better. Recently, [3] presented a two-step weighted LS estimator, yet it requires perfect knowledge of the variance of the shadowing effect. Moreover, in practice, if the PLE and anchor location information is inaccurate (e.g., especially in military scenarios, some critical information might be unreliable), uncertainties have to be considered into the constructed data model for DRSS-based localization. However, very few results exist in this area, even for RSS-based localization. In a nutshell, the research on DRSS-based localization is still in its infancy and requires more attention.

To enrich the research on DRSS-based localization and to tackle the earlier mentioned problems, the first contribution of this paper is to introduce a new whitened model for DRSS-based localization with different unknown transmit powers. Based on this model, an advanced best linear unbiased estimator (A-BLUE), a Lagrangian estimator (LE) and a robust SDP-based estimator (RSDPE) are respectively proposed, assuming the PLE is known, in which the RSDPE is particularly designed to cope with model uncertainties. Their computational complexities are discussed and verified by experiments. We also conduct simulations to study their performances under different noise conditions, different PLEs, imperfect PLE knowledge and inaccurate anchor location information. Finally, after accumulating enough insights by studying the three proposed estimators, we take a step further and develop an RSDP-based block coordinate descent estimator (RSDP-BCDE) to cope with the case when the PLE is totally unknown. Some issues related to a real-life implementation are also considered and discussed.

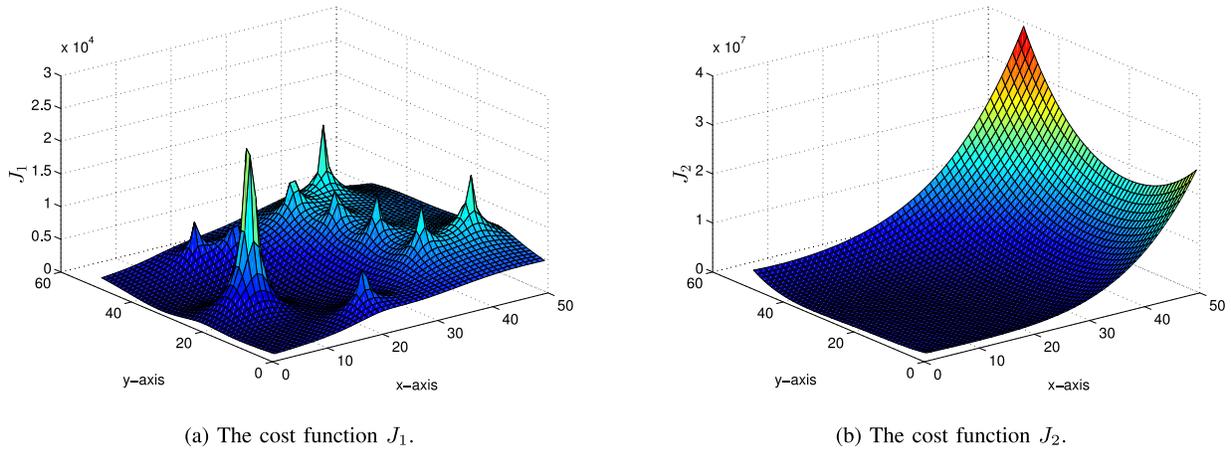

(a) The cost function $J_1$.   (b) The cost function $J_2$.

Fig. 1. Comparison between the least squares cost functions related to the models in (2) and (4): In $\mathbb{R}^2$, the target node is at $(28.7, 16.3)$; the anchor nodes are located at $(22.5, 10.2)$, $(44.9, 38.1)$, $(44.1, 14.2)$, $(33.6, 33.2)$, $(6.1, 20.3)$, $(13.7, 35.8)$, $(14.1, 44.8)$, $(41.3, 19.5)$, $(24.9, 34.7)$ and $(41.7, 30.5)$, of which the RN is selected as the one at $(22.5, 10.2)$. Obviously, the target node cannot overlap with the anchor nodes and hence every anchor location becomes a singular point in $J_1$ yielding multiple minima while $J_2$ has only a single optimal point.

After this brief introduction, Section II elaborates on our new whitened model for DRSS-based localization with different unknown transmit powers, which is used throughout this paper. Then, three different kinds of estimators based on a known PLE (i.e., the A-BLUE, the LE and the RSDPE) are proposed in Section III. Their complexities and performances in different situations are also analyzed and studied by numerical simulations. Based on those studies, Section IV presents a solution (i.e., the RSDP-BCDE) to the DRSS-based localization problem when the PLE is completely unknown. We also simulate and discuss this solution at the end of Section IV. Finally, Section V summarizes the results of this paper.

## II. WHITENED MODEL FOR DRSS-BASED LOCALIZATION

We would firstly like to cope with the non-linearity issue of (2). To do this, we transform (2) into

$$||\mathbf{x} - \mathbf{s}_i||_2^2 P'_{i,1} = \Delta P'_{0,i,1} \chi'_{i,1} ||\mathbf{x} - \mathbf{s}_1||_2^2, \ i \neq 1, \quad (3)$$

where $P'_{i,1} \triangleq 10^{\frac{P_{i,1}}{5\gamma}}$, $\Delta P'_{0,i,1} \triangleq 10^{\frac{\Delta P_{0,i,1}}{5\gamma}}$, and $\chi'_{i,1} \triangleq 10^{\frac{\chi_{i,1}}{5\gamma}}$. Then, unfolding the Euclidean norm in (3), introducing $d_1^2 \triangleq ||\mathbf{x} - \mathbf{s}_1||_2^2$ and stacking equations into matrices, our linear model for DRSS-based localization can be written as

$$\boldsymbol{p} = \boldsymbol{\Psi}\boldsymbol{\theta} + \boldsymbol{\epsilon}, \quad (4)$$

where

$$\boldsymbol{\Psi} \triangleq \begin{bmatrix} \vdots & \vdots \\ 2\mathbf{s}_1^T - 2P'_{i,1}\mathbf{s}_i^T & P'_{i,1} - 1 \\ \vdots & \vdots \end{bmatrix}, \ \boldsymbol{\theta} \triangleq [\mathbf{x}^T, ||\mathbf{x}||_2^2]^T,$$

$$\boldsymbol{p} \triangleq \begin{bmatrix} \vdots \\ ||\mathbf{s}_1||_2^2 - ||\mathbf{s}_i||_2^2 P'_{i,1} \\ \vdots \end{bmatrix}, \text{ and } \boldsymbol{\epsilon} \triangleq \begin{bmatrix} \vdots \\ d_1^2(1 - \Delta P'_{0,i,1}\chi'_{i,1}) \\ \vdots \end{bmatrix}.$$

By respectively denoting $[\cdot]_i$ as the $i$-th element of a vector and $[\cdot]_{1:i}$ as the subvector containing the first $i$ elements of a vector, we observe that $[\boldsymbol{\theta}]_{1:d}$ corresponds to the target location $\mathbf{x}$ and, more importantly, a new parameter is introduced at $[\boldsymbol{\theta}]_{d+1}$ which corresponds to $||\mathbf{x}||_2^2$. Our optimization problem w.r.t. $\boldsymbol{\theta}$ obviously becomes easier and any estimate of $\boldsymbol{\theta}$ leads to an estimate of $\mathbf{x}$, i.e., $\hat{\mathbf{x}} = [\hat{\boldsymbol{\theta}}]_{1:d}$.

To be more explicit, the model (4) is smoother than (2). To illustrate that, let us apply the least squares criterion to (2) and (4), leading to the respective cost functions in $\mathbf{x}$:

$$J_1 = \sum_{i=2}^{N} \left[ P_{i,1} + 10\gamma \log_{10}\left(\frac{||\mathbf{x} - \mathbf{s}_i||_2}{||\mathbf{x} - \mathbf{s}_1||_2}\right) \right]^2$$

and

$$J_2 = ||\boldsymbol{\Psi}\begin{bmatrix} \mathbf{x} \\ ||\mathbf{x}||_2^2 \end{bmatrix} - \boldsymbol{p}||_2^2.$$

As depicted in Fig. 1, $J_1$ has multiple minima while $J_2$ becomes convex w.r.t. $\mathbf{x}$ yielding only a single optimal point. Note that we explicitly take the dependence in $\boldsymbol{\theta}$ into account when formulating $J_2$. In other words, we assume

$$[\boldsymbol{\theta}]_{1:d}^T[\boldsymbol{\theta}]_{1:d} = [\boldsymbol{\theta}]_{d+1},$$

which also implies that $\boldsymbol{\theta}$ is bound to a non-convex set since the dependence in $\boldsymbol{\theta}$ is considered.

To obtain our whitened model for DRSS-based localization, let us denote an element of $\boldsymbol{\epsilon}$ as $\epsilon_i = d_1^2(1 - \Delta P'_{0,i,1}\chi'_{i,1})$, $i \neq 1$. For a sufficiently small shadowing effect and transmit power deviation, $\epsilon_i$ can be approximated by its first-order Taylor series expansion[1]

$$\epsilon_i = d_1^2\left(1 - 10^{\frac{\Delta P_{0,i,1} + \chi_{i,1}}{5\gamma}}\right) = C(\Delta P_{0,i,1} + \chi_{i,1}), \quad (5)$$

which is apparently zero-mean yet mutually correlated, where $C \triangleq -\frac{\ln(10)d_1^2}{5\gamma}$ is a scaling factor. For the record, when the shadowing effect or the transmit power deviation grows very large, the approximation in (5) might become inaccurate. Notice that $\epsilon_i$ is subject to the PLE $\gamma$ as well as the distance $d_1$ from the target to the RN. We cannot do much about the PLE, but it is clear that choosing a close RN will suppress the model error $\epsilon$. Therefore, in this paper, the RN is chosen as the anchor node that has the highest RSS, since that anchor node is most likely the one that is closest to the target node. Note that in a mobile scenario, the RN should be updated in time, but we will not consider that in this paper.

---
[1]$a^x = 1 + x\ln(a) + \cdots + \frac{(x\ln(a))^n}{n!} + \cdots, \ -\infty < x < \infty.$

For convenience, we respectively define the correlated DRSS measurement noise in (2) as $n_{i,1} \triangleq \Delta P_{0,i,1} + \chi_{i,1}$ and the independent measurement noise as $n_i \triangleq \Delta P_{0,i} + \chi_i$. Recalling that $\Delta P_{0,i,1} = \Delta P_{0,i} - \Delta P_{0,1}$ and $\chi_{i,1} = \chi_i - \chi_1$, we have $n_{i,1} = n_i - n_1$. Hence, from (5), the unwhitened model error $\epsilon$ can be approximated as

$$\epsilon = C\Gamma \mathbf{n},$$

where $\mathbf{n}$ stacks all independent DRSS measurement noise terms $n_i$ and

$$\Gamma \triangleq \begin{bmatrix} -\mathbf{1}_{(N-1)\times 1} & \mathbf{I}_{N-1} \end{bmatrix}_{(N-1)\times N}, \quad (6)$$

with $\mathbf{I}$ the identity matrix, $\mathbf{0}$ the zero matrix and $\mathbf{1}$ the all-one matrix (sizes are mentioned in the subscript if needed). In this paper, we assume that $\Delta P_{0,i}$ is a zero-mean Gaussian variable with variance $\sigma_{P_0}^2$. Therefore, we can obtain $\epsilon \sim \mathcal{N}(\mathbf{0}, \Sigma_\epsilon)$ and the covariance matrix of $\epsilon$ can be computed as

$$\Sigma_\epsilon = C^2(\sigma_{P_0}^2 + \sigma_\chi^2)\Gamma\Gamma^T = C^2 \sigma_n^2 \Gamma\Gamma^T,$$

where $\sigma_n^2 \triangleq \sigma_{P_0}^2 + \sigma_\chi^2$ is the variance of the independent measurement noise $n_i$ (or simply called measurement noise from now on), i.e., $\mathbf{n} \sim \mathcal{N}(\mathbf{0}, \sigma_n^2 \mathbf{I}_N)$.

Finally, from (4), we can obtain the whitened model as

$$\Sigma_\epsilon^{-1/2} \mathbf{p} = \Sigma_\epsilon^{-1/2} \Psi\theta + \Sigma_\epsilon^{-1/2}\epsilon \quad (7a)$$

$$\Rightarrow (\Gamma\Gamma^T)^{-1/2}\mathbf{p} = (\Gamma\Gamma^T)^{-1/2}\Psi\theta + (\Gamma\Gamma^T)^{-1/2}\epsilon \quad (7b)$$

$$\Rightarrow \rho = \Phi\theta + \mathbf{v} \quad (7c)$$

where $\rho \triangleq (\Gamma\Gamma^T)^{-1/2}\mathbf{p}$, $\Phi \triangleq (\Gamma\Gamma^T)^{-1/2}\Psi$ and $\mathbf{v} \triangleq (\Gamma\Gamma^T)^{-1/2}\epsilon$. Obviously, the model error $\mathbf{v}$ in (7c) is whitened, since its covariance matrix is a scaled identity, i.e., $\Sigma_\mathbf{v} = C^2 \sigma_n^2 \mathbf{I}_{N-1}$.

An important observation that we would like to make about our whitened DRSS-based data model is that no information is lost by taking differences of RSSs, since our model can be alternatively derived from a properly whitened RSS-based model after orthogonally projecting out the unknown average power $\bar{P}_0$ (see Appendix B for details). As a result, the choice of the RN has no effect on the performance of the localization accuracy.

### III. ESTIMATORS FOR KNOWN PATH-LOSS MODEL

In this section, we assume that the PLE $\gamma$ is known and our derivations start from an exactly known data model. Considering our whitened model (7c) and ignoring the dependence in $\theta$, it is possible to formulate the following unconstrained optimization problem

$$\min_{\theta} ||\Phi\theta - \rho||_2^2. \quad (8)$$

This leads to the unconstrained best linear unbiased estimator (U-BLUE) for $\mathbf{x}$, which can be presented as

$$\begin{aligned} \hat{\mathbf{x}}_{u-blue} &= \left[\hat{\theta}_{u-blue}\right]_{1:d} \\ &= \left[(\Phi^T \Sigma_\mathbf{v}^{-1} \Phi)^{-1}\Phi^T \Sigma_\mathbf{v}^{-1}\rho\right]_{1:d} \\ &= \left[(\Phi^T \Phi)^{-1}\Phi^T \rho\right]_{1:d}. \end{aligned} \quad (9)$$

Note that the unknowns $C$ and $\sigma_n^2$ are eliminated in this solution. Although there are other similar least squares (LS) solutions [15], [17], [18], none of them is the BLUE since their data models are still coloured. Here, the U-BLUE will not perform very well as we will illustrate later on. Hence, in this section, we introduce two alternative methods based on an exactly known data model and then take some model uncertainties into account, which finally leads to a robust estimator for DRSS-based localization. We conduct simulations to study their performances under different noise conditions, different PLEs, imperfect PLE knowledge and inaccurate anchor location information. Their complexities are also studied and numerical results are presented. We end this section by discussing some practical issues.

#### A. Advanced Best Linear Unbiased Estimator

To boost the performance of the U-BLUE, we will take the dependence in $\theta$ into account and hence our optimization problem has to be reformulated as

$$\min_{\theta} ||\Phi\theta - \rho||_2^2 \quad (10a)$$

$$\text{subject to } [\theta]_{1:d}^T [\theta]_{1:d} = [\theta]_{d+1}. \quad (10b)$$

The commonly known method to solve this problem indirectly is by constructing a new data model [3], [16], [33]–[36]. For instance, the new model can be given by

$$\mathbf{g} = \mathbf{Q}\mathbf{z} + \mathbf{m}, \quad (11)$$

where $\mathbf{g} \triangleq [[\hat{\theta}_{u-blue}]_1^2, \cdots, [\hat{\theta}_{u-blue}]_d^2, [\hat{\theta}_{u-blue}]_{d+1}]^T$, $\mathbf{Q} \triangleq [\mathbf{I}_d, \mathbf{1}_{d\times 1}]^T$, $\mathbf{z} \triangleq [[\mathbf{x}]_1^2, \cdots, [\mathbf{x}]_d^2]^T$ and

$$\mathbf{m} \triangleq \begin{bmatrix} [\hat{\theta}_{u-blue}]_1^2 - [\mathbf{x}]_1^2 \\ \vdots \\ [\hat{\theta}_{u-blue}]_d^2 - [\mathbf{x}]_d^2 \\ [\hat{\theta}_{u-blue}]_{d+1} - ||\mathbf{x}||_2^2 \end{bmatrix} \approx \begin{bmatrix} 2[\mathbf{x}]_1([\hat{\theta}_{u-blue}]_1 - [\mathbf{x}]_1) \\ \vdots \\ 2[\mathbf{x}]_d([\hat{\theta}_{u-blue}]_d - [\mathbf{x}]_d) \\ [\hat{\theta}_{u-blue}]_{d+1} - ||\mathbf{x}||_2^2 \end{bmatrix}. \quad (12)$$

Based on this model, the location estimate considering the constraint (10b) can be obtained as

$$\hat{\mathbf{x}} = [\text{sign}([\hat{\theta}_{u-blue}]_1)\sqrt{[\hat{\mathbf{z}}]_1}, \cdots, \text{sign}([\hat{\theta}_{u-blue}]_d)\sqrt{[\hat{\mathbf{z}}]_d}]^T,$$

where $\text{sign}(\cdot)$ indicates the signum function and $\hat{\mathbf{z}}$ is an estimate of $\mathbf{z}$. However, note that this method actually estimates the squared element of the target location $\mathbf{x}$ and the squaring procedure on $\hat{\theta}_{u-blue}$, which leads to the new observation vector $\mathbf{g}$, might exacerbates the estimation error in $\hat{\theta}_{u-blue}$.

Here, we propose an advanced best linear unbiased estimator (A-BLUE) to solve (10) directly, which fine-tunes $\hat{\theta}_{u-blue}$ without any squaring procedure. Recalling from (9) that $\hat{\boldsymbol{\theta}}_{u-blue} = (\boldsymbol{\Phi}^T\boldsymbol{\Phi})^{-1}\boldsymbol{\Phi}^T\boldsymbol{\rho}$, the cost function in (10a) can be rewritten as

$$J = (\boldsymbol{\Phi}\boldsymbol{\theta} - \boldsymbol{\rho})^T(\boldsymbol{\Phi}\boldsymbol{\theta} - \boldsymbol{\rho})$$
$$= (\boldsymbol{\theta} - \hat{\boldsymbol{\theta}}_{u-blue})^T \boldsymbol{\Phi}^T\boldsymbol{\Phi}(\boldsymbol{\theta} - \hat{\boldsymbol{\theta}}_{u-blue}). \qquad (13)$$

In order to take the constraint (10b) into account, $\boldsymbol{\theta}$ has to be reformulated as a function of $[\boldsymbol{\theta}]_{1:d}$, i.e.,

$$\boldsymbol{\theta} = \begin{bmatrix} [\boldsymbol{\theta}]_{1:d} \\ [\boldsymbol{\theta}]_{1:d}^T [\boldsymbol{\theta}]_{1:d} \end{bmatrix}. \qquad (14)$$

By now using the first-order Taylor series expansion of $\boldsymbol{\theta} - \hat{\boldsymbol{\theta}}_{u-blue}$ for $[\boldsymbol{\theta}]_{1:d}$ in the vicinity of $\hat{\mathbf{x}}_{u-blue}$, we obtain

$$\boldsymbol{\theta} - \hat{\boldsymbol{\theta}}_{u-blue} = \boldsymbol{\theta}|_{[\boldsymbol{\theta}]_{1:d} = \hat{\mathbf{x}}_{u-blue}} - \hat{\boldsymbol{\theta}}_{u-blue}$$
$$+ \left. \frac{\partial \boldsymbol{\theta}}{\partial [\boldsymbol{\theta}]_{1:d}^T} \right|_{[\boldsymbol{\theta}]_{1:d} = \hat{\mathbf{x}}_{u-blue}} ([\boldsymbol{\theta}]_{1:d} - \hat{\mathbf{x}}_{u-blue})$$
$$= \boldsymbol{\tau} + \mathbf{G}([\boldsymbol{\theta}]_{1:d} - \hat{\mathbf{x}}_{u-blue}), \qquad (15)$$

where

$$\boldsymbol{\tau} \triangleq \boldsymbol{\theta}|_{[\boldsymbol{\theta}]_{1:d} = \hat{\mathbf{x}}_{u-blue}} - \hat{\boldsymbol{\theta}}_{u-blue} = \begin{bmatrix} \mathbf{0}_{d \times 1} \\ ||\hat{\mathbf{x}}_{u-blue}||_2^2 - [\hat{\boldsymbol{\theta}}_{u-blue}]_{d+1} \end{bmatrix}$$

and $\mathbf{G} \triangleq \left. \frac{\partial \boldsymbol{\theta}}{\partial [\boldsymbol{\theta}]_{1:d}^T} \right|_{[\boldsymbol{\theta}]_{1:d} = \hat{\mathbf{x}}_{u-blue}} = \begin{bmatrix} \mathbf{I}_d \\ 2\hat{\mathbf{x}}_{u-blue}^T \end{bmatrix}$.

Substituting (15) into (13), we obtain

$$J = (\boldsymbol{\tau} + \mathbf{G}([\boldsymbol{\theta}]_{1:d} - \hat{\mathbf{x}}_{u-blue}))^T \boldsymbol{\Phi}^T\boldsymbol{\Phi}(\boldsymbol{\tau}$$
$$+ \mathbf{G}([\boldsymbol{\theta}]_{1:d} - \hat{\mathbf{x}}_{u-blue})). \qquad (16)$$

Taking the derivative of (16) w.r.t. $[\boldsymbol{\theta}]_{1:d}$, we have

$$\frac{\partial J}{\partial [\boldsymbol{\theta}]_{1:d}} = 2\mathbf{G}^T\boldsymbol{\Phi}^T\boldsymbol{\Phi}\mathbf{G}([\boldsymbol{\theta}]_{1:d} - \hat{\mathbf{x}}_{u-blue}) + 2\mathbf{G}^T\boldsymbol{\Phi}^T\boldsymbol{\Phi}\boldsymbol{\tau}. \qquad (17)$$

Finally, by forcing (17) to $\mathbf{0}$, the A-BLUE can be expressed as

$$\hat{\mathbf{x}}_{a-blue} = \hat{\mathbf{x}}_{u-blue} - (\mathbf{G}^T\boldsymbol{\Phi}^T\boldsymbol{\Phi}\mathbf{G})^{-1}\mathbf{G}^T\boldsymbol{\Phi}^T\boldsymbol{\Phi}\boldsymbol{\tau}. \qquad (18)$$

### B. Lagrangian Estimator

The A-BLUE approximates (15) by linearizing it around $[\boldsymbol{\theta}]_{1:d} = \hat{\mathbf{x}}_{u-blue}$, which implies that its accuracy will certainly be degraded if there is a large estimation error in the U-BLUE. In this subsection, we would like to go one step further to find an estimator without any approximation.

In order to do so, we need to rewrite the constraint in (10b) and reformulate our optimization problem (10) as

$$\min_{\boldsymbol{\theta}} ||\boldsymbol{\Phi}\boldsymbol{\theta} - \boldsymbol{\rho}||_2^2 \qquad (19a)$$

$$\text{subject to } \boldsymbol{\theta}^T\mathbf{A}\boldsymbol{\theta} + 2\mathbf{b}^T\boldsymbol{\theta} = 0, \qquad (19b)$$

where $\mathbf{A} \triangleq \begin{bmatrix} \mathbf{I}_d & \mathbf{0} \\ \mathbf{0} & 0 \end{bmatrix}$ and $\mathbf{b} \triangleq \begin{bmatrix} \mathbf{0}_{d \times 1} \\ -\frac{1}{2} \end{bmatrix}$. The Lagrangian of (19) is

$$L(\boldsymbol{\theta}; \lambda) = (\boldsymbol{\Phi}\boldsymbol{\theta} - \boldsymbol{\rho})^T(\boldsymbol{\Phi}\boldsymbol{\theta} - \boldsymbol{\rho}) + \lambda(\boldsymbol{\theta}^T\mathbf{A}\boldsymbol{\theta} + 2\mathbf{b}^T\boldsymbol{\theta}), \qquad (20)$$

where $\lambda$ is the Lagrangian multiplier. Taking the derivative of (20) w.r.t. $\boldsymbol{\theta}$, we have

$$\frac{\partial L(\boldsymbol{\theta}; \lambda)}{\partial \boldsymbol{\theta}} = 2\boldsymbol{\Phi}^T\boldsymbol{\Phi}\boldsymbol{\theta} - 2\boldsymbol{\Phi}\boldsymbol{\rho} + 2\lambda\mathbf{A}\boldsymbol{\theta} + 2\lambda\mathbf{b} \qquad (21)$$

and forcing (21) to 0 leads to our Lagrangian estimator (LE) which is given by

$$\hat{\boldsymbol{\theta}}_{le}(\lambda) = (\boldsymbol{\Phi}^T\boldsymbol{\Phi} + \lambda\mathbf{A})^{-1}(\boldsymbol{\Phi}^T\boldsymbol{\rho} + \lambda\mathbf{b}). \qquad (22)$$

Since $\lambda$ is unknown, it is required to find an appropriate value for $\lambda$. A similar problem also appears in [37], [38], where all possible values of $\lambda$ should be calculated to determine the desired one. Note that some of those values might lead to a maximum of the Lagrangian in (20), since the second-order optimality conditions are not examined [9]. Besides, the above method is very cumbersome and, recalling the fact that $\boldsymbol{\theta}$ is bound to a nonconvex set, a suboptimal value of $\lambda$ might be selected, yielding a local solution. Without going into many details, we will not further discuss it. Here, the idea is to firstly pinpoint an interval for $\lambda$, in which only one single global solution is guaranteed, and then to search for that solution.

To find such an interval, note that the solution in (22) is a minimum of the Lagrangian in (20) if the Hessian of (20) is positive semidefinite, i.e.,

$$\boldsymbol{\Phi}^T\boldsymbol{\Phi} + \lambda\mathbf{A} \succeq 0$$
$$\Rightarrow (\boldsymbol{\Phi}^T\boldsymbol{\Phi})^{\frac{1}{2}} (\mathbf{I}_{N-1} \qquad \qquad . \qquad (23)$$
$$+ \lambda(\boldsymbol{\Phi}^T\boldsymbol{\Phi})^{-\frac{1}{2}}\mathbf{A}(\boldsymbol{\Phi}^T\boldsymbol{\Phi})^{-\frac{1}{2}})(\boldsymbol{\Phi}^T\boldsymbol{\Phi})^{\frac{1}{2}} \succeq 0$$

In order to guarantee (23), the eigenvalues of $\mathbf{I}_{N-1} + \lambda(\boldsymbol{\Phi}^T\boldsymbol{\Phi})^{-\frac{1}{2}}\mathbf{A}(\boldsymbol{\Phi}^T\boldsymbol{\Phi})^{-\frac{1}{2}}$ should be all non-negative. Obviously, all the eigenvalues of $(\boldsymbol{\Phi}^T\boldsymbol{\Phi})^{-\frac{1}{2}}\mathbf{A}(\boldsymbol{\Phi}^T\boldsymbol{\Phi})^{-\frac{1}{2}}$ are non-negative. Then denoting the largest eigenvalue of $(\boldsymbol{\Phi}^T\boldsymbol{\Phi})^{-\frac{1}{2}}\mathbf{A}(\boldsymbol{\Phi}^T\boldsymbol{\Phi})^{-\frac{1}{2}}$ as $\lambda_{max}$, we need $1 + \lambda\lambda_{max} \geq 0$, which provides a useful interval for $\lambda$ as

$$\mathcal{I} = (-1/\lambda_{max}, \infty).$$

On such an interval, we can find the desired value of $\lambda$, say $\hat{\lambda}_{le}$, such that

$$\hat{\boldsymbol{\theta}}_{le}(\hat{\lambda}_{le})^T \mathbf{A}\hat{\boldsymbol{\theta}}_{le}(\hat{\lambda}_{le}) + 2\mathbf{b}^T\hat{\boldsymbol{\theta}}_{le}(\hat{\lambda}_{le}) = 0. \qquad (24)$$

Then, the Lagrangian estimator (LE) for $\mathbf{x}$ can be obtained as $\hat{\mathbf{x}}_{le} = [\hat{\boldsymbol{\theta}}_{le}(\hat{\lambda}_{le})]_{1:d}$.

Now the problems left are how to search for $\hat{\lambda}_{le}$ and whether or not the LE yields the global solution. Before going into the details, it is important to firstly realize that the problem (19) is a quadratically constrained quadratic program (QCQP) which can be cast as a generalized trust region subproblem (GTRS) [39], for which an optimal solution can be found within a bounded interval, i.e., the interval $\mathcal{I}$. In this paper, we actually consider a simpler case with an equality constraint (19b) rather than an inequality constraint, yet some results can still be used to support the following discussions.

To search for $\hat{\lambda}_{le}$, let us define a function $f(\lambda)$ as $f(\lambda) \triangleq \hat{\boldsymbol{\theta}}_{le}(\lambda)^T \mathbf{A}\hat{\boldsymbol{\theta}}_{le}(\lambda) + 2\mathbf{b}^T\hat{\boldsymbol{\theta}}_{le}(\lambda)$, which is already known to be strictly decreasing on the interval $\mathcal{I}$ [40, Theorem 5.2], such that $\hat{\lambda}_{le}$, which satisfies the constraint (24), can be effectively found by the bisection method. Next, the LE is guaranteed as a global solution [40, Theorem 3.2], since it follows the *Karush-Kuhn-Tucker* (KKT) conditions. This also indicates that there only exists one solution, i.e., the global solution, in the interval $\mathcal{I}$, which is the reason why it is called the trust region. Besides, note that the case $\hat{\lambda}_{le} = -1/\lambda_{max}$ is called the *hard case* [41] (since it is relatively difficult to solve), which is very rare and has never been seen in our numerous simulations. The *hard case* is also found to be very rare in other papers, e.g., in [31], [42].

### C. Robust Semidefinite Programming Based Estimator

The previously proposed estimators are both based on an exactly known data model. However, when the data model is not perfectly known due to an imperfect PLE estimate or inaccurate anchor location information, a huge bias will obviously occur in these estimates. Therefore, in this subsection, we present a robust semidefinite programming based estimator (RSDPE) that can cope with such model uncertainties.

First, after using the *Schur complement* [43] and forming some linear matrix inequalities (LMIs), we equivalently rewrite the constraint in (10b) as

$$\begin{bmatrix} \mathbf{I}_d & [\boldsymbol{\theta}]_{1:d} \\ [\boldsymbol{\theta}]_{1:d}^T & [\boldsymbol{\theta}]_{d+1} \end{bmatrix} \succeq \mathbf{0}, \qquad (25a)$$

$$\text{rank}\left( \begin{bmatrix} \mathbf{I}_d & [\boldsymbol{\theta}]_{1:d} \\ [\boldsymbol{\theta}]_{1:d}^T & [\boldsymbol{\theta}]_{d+1} \end{bmatrix} \right) = d. \qquad (25b)$$

The semidefinite relaxation (SDR) approach then relaxes the set of $\boldsymbol{\theta}$ by dropping the rank constraint in (25b). This procedure is also used in [15], [19]–[23], but they all assume an exactly known data model.

We want to go one step further and consider an uncertain $\boldsymbol{\Phi}$ as $\boldsymbol{\Phi}^\circ \triangleq \boldsymbol{\Phi} + \boldsymbol{\Delta}_{\boldsymbol{\Phi}}$, where the perturbation matrix $\boldsymbol{\Delta}_{\boldsymbol{\Phi}}$ collects the uncertainties caused by an imperfect PLE estimate or inaccurate anchor location information. Although the data model is not exactly known, a known upper bound $\zeta$ for $||\boldsymbol{\Delta}_{\boldsymbol{\Phi}}||_2$ could be very helpful, i.e., $||\boldsymbol{\Delta}_{\boldsymbol{\Phi}}||_2 \leq \zeta$, where $||\cdot||_2$ here denotes the spectral norm, i.e., the largest singular value of the corresponding matrix.

The idea of the RSDPE is to cope with the worst-case model uncertainties using the SDP procedure. Therefore, we reformulate (10) as a minmax SDP optimization problem

$$\min_{\boldsymbol{\theta}, t} \max_{||\boldsymbol{\Delta}_{\boldsymbol{\Phi}}||_2 \leq \zeta} t \qquad (26a)$$

$$\text{subject to} \begin{bmatrix} \mathbf{I}_{N-1} & (\boldsymbol{\Phi}^\circ - \boldsymbol{\Delta}_{\boldsymbol{\Phi}})\boldsymbol{\theta} - \boldsymbol{\rho} \\ ((\boldsymbol{\Phi}^\circ - \boldsymbol{\Delta}_{\boldsymbol{\Phi}})\boldsymbol{\theta} - \boldsymbol{\rho})^T & t \end{bmatrix} \succeq \mathbf{0}, \qquad (26b)$$

$$\begin{bmatrix} \mathbf{I}_d & [\boldsymbol{\theta}]_{1:d} \\ [\boldsymbol{\theta}]_{1:d}^T & [\boldsymbol{\theta}]_{d+1} \end{bmatrix} \succeq \mathbf{0}. \qquad (26c)$$

where $t$ is an auxiliary slack variable.

Note that $\boldsymbol{\Delta}_{\boldsymbol{\Phi}}$ only affects the constraint (26b) and hence we can isolate $\boldsymbol{\Delta}_{\boldsymbol{\Phi}}$ in (26b) as

$$\mathbf{B}(\boldsymbol{\theta}, t) \succeq \begin{bmatrix} \mathbf{0} & \boldsymbol{\Delta}_{\boldsymbol{\Phi}}\boldsymbol{\theta} \\ \mathbf{0} & 0 \end{bmatrix} + \begin{bmatrix} \mathbf{0} & \mathbf{0} \\ (\boldsymbol{\Delta}_{\boldsymbol{\Phi}}\boldsymbol{\theta})^T & 0 \end{bmatrix}$$

$$\Rightarrow \mathbf{B}(\boldsymbol{\theta}, t) \succeq \mathbf{T}^T \boldsymbol{\Delta}_{\boldsymbol{\Phi}} \mathbf{L}(\boldsymbol{\theta}) + \mathbf{L}(\boldsymbol{\theta})^T \boldsymbol{\Delta}_{\boldsymbol{\Phi}}^T \mathbf{T}, \qquad (27)$$

where

$$\mathbf{B}(\boldsymbol{\theta}, t) \triangleq \begin{bmatrix} \mathbf{I}_{N-1} & (\boldsymbol{\Phi}^\circ \boldsymbol{\theta} - \boldsymbol{\rho}) \\ (\boldsymbol{\Phi}^\circ \boldsymbol{\theta} - \boldsymbol{\rho})^T & t \end{bmatrix},$$

$\mathbf{T} \triangleq [\mathbf{I}_{N-1} \quad \mathbf{0}]$ and $\mathbf{L}(\boldsymbol{\theta}) \triangleq [\mathbf{0} \quad \boldsymbol{\theta}]$. Obviously, for the maximization in (26), the constraint (27) has to be reformulated considering the worst-case $\boldsymbol{\Delta}_{\boldsymbol{\Phi}}$.

To do so, we can easily state that

$$\mathbf{B}(\boldsymbol{\theta}, t) \succeq \mathbf{T}^T \boldsymbol{\Delta}_{\boldsymbol{\Phi}} \mathbf{L}(\boldsymbol{\theta}) + \mathbf{L}(\boldsymbol{\theta})^T \boldsymbol{\Delta}_{\boldsymbol{\Phi}}^T \mathbf{T}, \forall \boldsymbol{\Delta}_{\boldsymbol{\Phi}} : ||\boldsymbol{\Delta}_{\boldsymbol{\Phi}}||_2 \leq \zeta \qquad (28)$$

if and only if

$$\check{\mathbf{x}}^T \mathbf{B}(\boldsymbol{\theta}, t)\check{\mathbf{x}} \geq \max_{||\boldsymbol{\Delta}_{\boldsymbol{\Phi}}||_2 \leq \zeta} \{\check{\mathbf{x}}^T \mathbf{T}^T \boldsymbol{\Delta}_{\boldsymbol{\Phi}} \mathbf{L}(\boldsymbol{\theta})\check{\mathbf{x}} + \check{\mathbf{x}}^T \mathbf{L}(\boldsymbol{\theta})^T \boldsymbol{\Delta}_{\boldsymbol{\Phi}}^T \mathbf{T}\check{\mathbf{x}}\}$$

$$= \max_{||\boldsymbol{\Delta}_{\boldsymbol{\Phi}}||_2 \leq \zeta} \{2||\boldsymbol{\Delta}_{\boldsymbol{\Phi}} \mathbf{L}(\boldsymbol{\theta})\check{\mathbf{x}}||_2 \ ||\mathbf{T}\check{\mathbf{x}}||_2\}$$

$$= 2\zeta ||\mathbf{L}(\boldsymbol{\theta})\check{\mathbf{x}}||_2 ||\mathbf{T}\check{\mathbf{x}}||_2, \forall \check{\mathbf{x}} \in \mathbb{R}^N. \qquad (29)$$

After introducing the bound $\zeta$ into (29), a new problem arises since the vector $\mathbf{T}\check{\mathbf{x}} \in \mathbb{R}^{N-1}$ does not have the same size as the vector $\mathbf{L}(\boldsymbol{\theta})\check{\mathbf{x}} \in \mathbb{R}^{d+1}$. To bypass this issue, we introduce a new auxiliary vector $\check{\mathbf{y}} \in \mathbb{R}^{d+1}$, which is bounded using $\check{\mathbf{x}}$, such that we can use the Cauchy-Schwarz inequality on (29) to unfold the norm. To be specific, only after the worst-case constraint (29) is reformulated as

$$\check{\mathbf{x}}^T \mathbf{B}(\boldsymbol{\theta}, t)\check{\mathbf{x}} \geq 2\zeta ||\check{\mathbf{y}}||_2 ||\mathbf{L}(\boldsymbol{\theta})\check{\mathbf{x}}||_2, \forall \check{\mathbf{x}}, \check{\mathbf{y}} : ||\mathbf{T}\check{\mathbf{x}}||_2 \geq ||\check{\mathbf{y}}||_2, \qquad (30)$$

we can obtain a new constraint without the norm from (30) as

$$\check{\mathbf{x}}^T \mathbf{B}(\boldsymbol{\theta}, t)\check{\mathbf{x}} \geq \zeta(\check{\mathbf{y}}^T \mathbf{L}(\boldsymbol{\theta})\check{\mathbf{x}} + \check{\mathbf{x}}^T \mathbf{L}(\boldsymbol{\theta})^T \check{\mathbf{y}}), \forall \check{\mathbf{x}}, \check{\mathbf{y}} : ||\mathbf{T}\check{\mathbf{x}}||_2$$
$$\geq ||\check{\mathbf{y}}||_2. \qquad (31)$$

Although both (30) and (31) consider the worst-case $\boldsymbol{\Delta}_{\boldsymbol{\Phi}}$, we have to use the latter one to facilitate the derivations, which is actually a weaker condition due to the Cauchy-Schwarz inequality. Then, for convenience, we respectively rewrite $||\mathbf{T}\check{\mathbf{x}}||_2 \geq ||\check{\mathbf{y}}||_2$ as

$$\begin{bmatrix} \check{\mathbf{x}} \\ \check{\mathbf{y}} \end{bmatrix}^T \begin{bmatrix} \mathbf{T}^T\mathbf{T} & \mathbf{0} \\ \mathbf{0} & -\mathbf{I}_{d+1} \end{bmatrix} \begin{bmatrix} \check{\mathbf{x}} \\ \check{\mathbf{y}} \end{bmatrix} \geq 0 \qquad (32)$$

and (31) as

$$\begin{bmatrix} \check{\mathbf{x}} \\ \check{\mathbf{y}} \end{bmatrix}^T \begin{bmatrix} \mathbf{B}(\boldsymbol{\theta}, t) & -\zeta\mathbf{L}(\boldsymbol{\theta})^T \\ -\zeta\mathbf{L}(\boldsymbol{\theta}) & \mathbf{0} \end{bmatrix} \begin{bmatrix} \check{\mathbf{x}} \\ \check{\mathbf{y}} \end{bmatrix} \geq 0, \qquad (33)$$

where note that (33) is a necessary condition to (32).

Finally, according to the $S$-procedure [44, p. 23], the implication that (32) leads to (33) holds true if and only if there exists

an $\alpha$ such that

$$\begin{bmatrix} \mathbf{B}(\boldsymbol{\theta},t) & -\zeta \mathbf{L}(\boldsymbol{\theta})^T \\ -\zeta \mathbf{L}(\boldsymbol{\theta}) & \mathbf{0} \end{bmatrix} - \alpha \begin{bmatrix} \mathbf{T}^T \mathbf{T} & \mathbf{0} \\ \mathbf{0} & -\mathbf{I}_{d+1} \end{bmatrix} \succeq \mathbf{0} \quad (34a)$$

$$\Leftrightarrow \begin{bmatrix} (1-\alpha)\mathbf{I}_{N-1} & \boldsymbol{\Phi}^\circ \boldsymbol{\theta} - \boldsymbol{\rho} & \mathbf{0} \\ (\boldsymbol{\Phi}^\circ \boldsymbol{\theta} - \boldsymbol{\rho})^T & t & -\zeta \boldsymbol{\theta}^T \\ \mathbf{0} & -\zeta \boldsymbol{\theta} & \alpha \mathbf{I}_{d+1} \end{bmatrix} \succeq \mathbf{0}. \quad (34b)$$

Replacing (26b) by the new constraint (34b) leads to the following SDP optimization problem

$$\min_{\boldsymbol{\theta},t,\alpha} t \quad (35a)$$

$$\text{subject to} \begin{bmatrix} (1-\alpha)\mathbf{I}_{N-1} & \boldsymbol{\Phi}^\circ \boldsymbol{\theta} - \boldsymbol{\rho} & \mathbf{0} \\ (\boldsymbol{\Phi}^\circ \boldsymbol{\theta} - \boldsymbol{\rho})^T & t & -\zeta \boldsymbol{\theta}^T \\ \mathbf{0} & -\zeta \boldsymbol{\theta} & \alpha \mathbf{I}_{d+1} \end{bmatrix} \succeq \mathbf{0}, \quad (35b)$$

$$\begin{bmatrix} \mathbf{I}_d & [\boldsymbol{\theta}]_{1:d} \\ [\boldsymbol{\theta}]_{1:d}^T & [\boldsymbol{\theta}]_{d+1} \end{bmatrix} \succeq \mathbf{0}, \quad (35c)$$

which can be solved by *CVX* [45], [46]. The solution is our new RSDPE.

To end this subsection, we discuss how to determine the value of $\zeta$. One possibility is that $\zeta$ can be computed from the total least squares (TLS) method [47]. More specifically, we can compute the singular value decomposition (SVD) of the augmented matrix $[\boldsymbol{\Phi}^\circ \ \boldsymbol{\rho}] = \mathbf{U}\boldsymbol{\Sigma}\mathbf{V}^T$ and the corrected $[\boldsymbol{\Phi}^\circ \ \boldsymbol{\rho}]$ is given by $[\hat{\boldsymbol{\Phi}} \ \hat{\boldsymbol{\rho}}] = \mathbf{U}\hat{\boldsymbol{\Sigma}}\mathbf{V}^T$, where $\hat{\boldsymbol{\Sigma}}$ is obtained by forcing the $(d+2)$-th diagonal of $\boldsymbol{\Sigma}$ to 0, which is the typical low-rank approximation [48]. In fact, $\hat{\boldsymbol{\Phi}}$ can be viewed as an estimate of the exact $\boldsymbol{\Phi}$ and also observe that $||\boldsymbol{\Delta}_{\boldsymbol{\Phi}}||_2 = ||\boldsymbol{\Phi}^\circ - \boldsymbol{\Phi}||_2 \leq ||\boldsymbol{\Phi}^\circ - \boldsymbol{\Phi}||_F$, where $||\cdot||_F$ indicates the Frobenius norm or the Hilbert-Schmidt norm. Therefore, in this paper, $\zeta$ is computed as $\zeta = ||\boldsymbol{\Phi}^\circ - \hat{\boldsymbol{\Phi}}||_F$, which will also be used in our simulations later.

### D. Complexity Analysis

We now calculate the computational complexity of the different methods without considering the whitening procedure [49]. It is easy to derive that the complexity of the U-BLUE is $O[d^2 N]$. As for the A-BLUE, its complexity is $O[d^2 N^2]$ considering that the extra cost is mainly comes from the second step in (18).

For the LE, the complexity is mostly due to the bisection method. Suppose that the bisection method takes $K$ steps to find an appropriate $\lambda$, which has already been observed to be around 20. In each iteration, first the $\hat{\boldsymbol{\theta}}_{le}(\lambda)$ in (22) is computed and then $f(\lambda)$ is calculated to check if the outcome is smaller than the tolerance. As a result, the cost of each iteration is $O[d^2 N^2]$ and hence the complexity of the LE is $O[K d^2 N^2]$.

Finally, let us focus on the RSDPE. We consider the *worst-case* complexity for solving (35), which can be derived from employing the interior-point algorithm [9]. This implies that the complexity for each iteration is $O[d^2 N^2]$ and the iteration number is bounded by $O[\sqrt{N} ln(1/\xi)]$ [43], where $\xi$ is the iteration tolerance. Therefore, the complexity of the RSDPE in this paper is $O[d^2 N^{2.5} ln(1/\xi)]$.

Obviously, the RSDPE has the highest complexity among all the proposed estimators. To verify the complexities, we conduct an experiment in a 2-D space with 10 anchor nodes and use the average computational time as a complexity measure. The experiment is implemented in Matlab R2013b on a Lenovo IdeaPad Y570 (Processor 2.0 GHz Intel Core i7, Memory 8 GB). We observe that the U-BLUE and the A-BLUE respectively have the least and the second least average computational time of 0.026 ms and 0.049 ms while the RSDPE holds the highest one with 314.0 ms. Compared with the others, the complexity of the LE is reasonable with a computational time of 4.8 ms.

### E. Numerical Results

We have conducted a Monte Carlo (MC) simulation using 1000 trials on a 50 m $\times$ 50 m field, where one target node is randomly deployed for each trial. Our proposed estimators are compared against two existing methods: the RSS-based joint estimator (SDP-RSS) which is the best estimator from [15] and applies the SDP procedure on a $\ell_1$-norm approximation to jointly estimate the transmit power and the target location; and the recent DRSS-based two-step weighted least squares estimator (WLS-DRSS) of [3] which requires perfect knowledge of the variance of the measurement noise $\sigma_n^2$. Recall that, in this paper, the measurement noise includes the shadowing effect and transmit power derivations. For computing the *Cramér-Rao lower bound* (CRLB), see Appendix C. The root mean square error (RMSE) is used to evaluate the performance of all estimators.

*1) Impact of the Anchor Node Placement:* We first discuss the impact of the anchor node placement, where the simulation is conducted with a perfectly known data model. Obviously, a good anchor node placement is very significant for any kind of localization. To be specific, if the anchor nodes get clustered, the measurements and the locations of those anchor nodes are both very close to each other, which easily leads to an ill-posed optimization problem for localization. For example, the cluster of anchor nodes in Fig. 2(d) causes the matrix $\boldsymbol{\Phi}$ to be ill-conditioned, thus making our optimization problem very susceptible to the measurement noise. To verify that, two simulations are conducted one with a good and one with a bad anchor node placement. These two anchor node placements and the numerical results are shown in Fig. 2. Clearly, when a good anchor node placement is considered, our proposed estimators, especially the LE, can yield a performance very close to the CRLB with a known PLE, i.e., $CRLB_3$ in Appendix C. However, a bad anchor node placement causes a considerable gap between our proposed estimators and $CRLB_3$.

To include the effect of different anchor node placements, in the following simulations, 10 anchor nodes will be randomly deployed within the 50 m $\times$ 50 m field in each simulation trial and, hence, an average CRLB will be considered since the CRLB varies over the anchor node placement.

*2) Impact of the Measurement Noise:* In Fig. 3, we study all estimators with a perfectly known data model under large and small measurement noise when the PLE is known and fixed at $\gamma = 4$. The following observations can be made:
  i) *U-BLUE:* Even based on a whitened data model and being a BLUE, the U-BLUE still yields a very bad performance

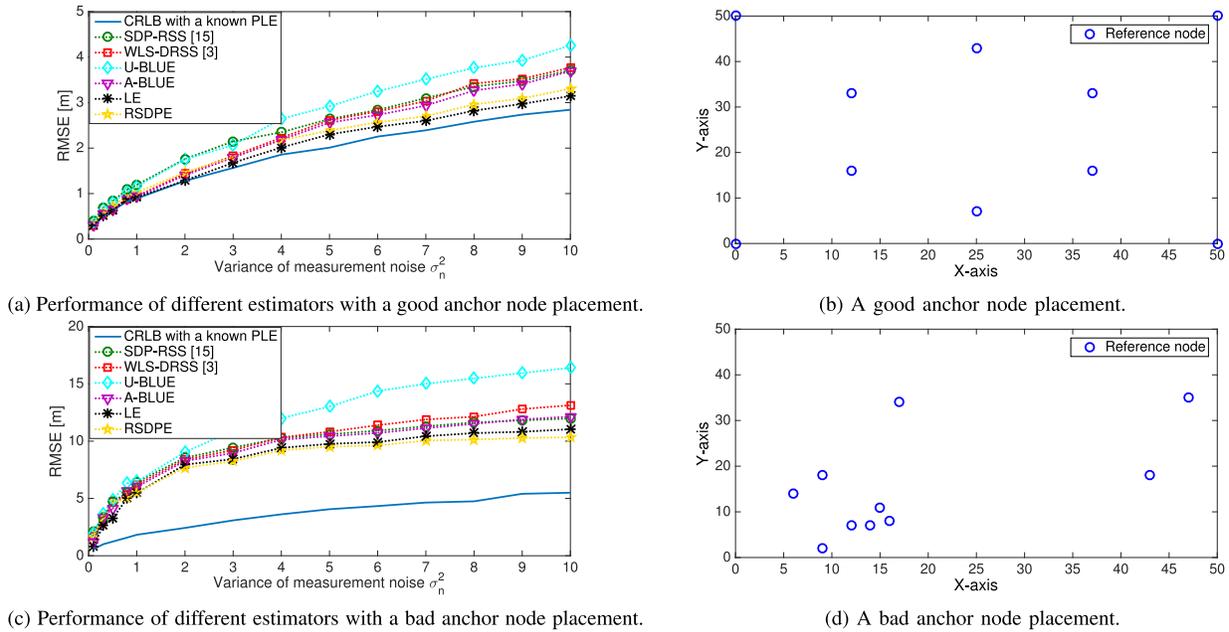

(a) Performance of different estimators with a good anchor node placement.

(b) A good anchor node placement.

(c) Performance of different estimators with a bad anchor node placement.

(d) A bad anchor node placement.

Fig. 2. Impact of the anchor node placement: In $\mathbb{R}^2$, 10 anchor nodes are considered with different placements and the target node is randomly deployed within a 50 m × 50 m field where the path-loss exponent is considered $\gamma = 4$. The anchor location information is accurate.

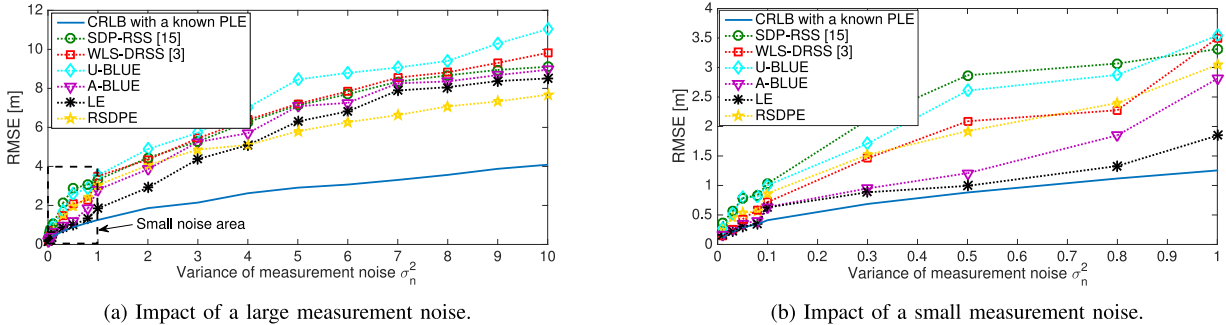

(a) Impact of a large measurement noise.

(b) Impact of a small measurement noise.

Fig. 3. Performance comparison of different estimators under different noise conditions when the actual PLE is known and fixed at $\gamma = 4$. The anchor location information is accurate.

especially under a large measurement noise since it does not consider the dependence in the parameter vector $\boldsymbol{\theta}$.

ii) *WLS-DRSS:* Even requiring perfect knowledge of the variance of the measurement noise $\sigma_n^2$ to construct its system model and its weighting matrices, the WLS-DRSS is still no better than any of our proposed DRSS-based estimators except for the U-BLUE. This is because many approximations are used in its derivation and the DRSS measurements are used themselves to construct the weighting matrices. Therefore, when the measurement noise grows more severe, those approximations become more inaccurate and the DRSS measurements are more corrupted, making the weighting matrices less effective as they are in a small noise situation.

iii) *A-BLUE:* Even without any knowledge of the variance of the measurement noise, the A-BLUE still has a better performance than the WLS-DRSS under a large measurement noise, as shown in Fig. 3(a). Under a small measurement noise, the A-BLUE becomes very accurate and only worse than the LE, as shown in Fig. 3(b). To explain this, the approximation in the second step of the A-BLUE is taken in the vicinity of the estimate $\hat{\mathbf{x}}_{u-blue}$ from the U-BLUE, which remains accurate under a small measurement noise. However, under a severe measurement noise, the U-BLUE yields a very bad performance and hence it becomes more difficult for the A-BLUE to fine-tune the U-BLUE estimate.

iv) *LE:* The LE outperforms all the other estimators under a small measurement noise due to the fact that it requires neither any approximation nor dropping a constraint. In fact, the LE is the exact solution to our optimization problem in (19) if our model error $\boldsymbol{v}$ is perfectly whitened. Therefore, we can observe from Fig. 3(b) that its performance is very close to the CRLB. However, the LE becomes only the second best estimator under a large measurement noise. To explain that, we need to recall that the approximation in (5) might become inaccurate under a large measurement noise, thus making our proposed whitened model not as effective as it is under a small measurement noise. On the other hand, a bad anchor node placement can also exacerbate the impact of a large measurement noise, which heavily deteriorates the performance of our proposed estimators.

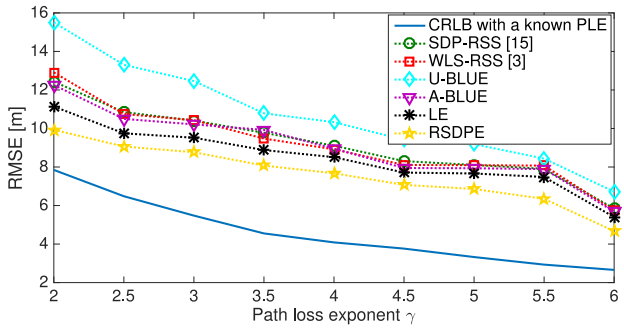

Fig. 4. Performance of different estimators under different PLEs when the variance of the measurement noise is $\sigma_n^2 = 10$. The anchor location information is accurate.

v) *RSDPE:* Unlike the A-BLUE, the RSDPE does not use any approximation to deal with the non-linearity issue. Instead, the RSDPE uses the SDR procedure in (26) to guarantee a global yet suboptimal solution at a price of dropping the rank constraint in (25b). This explains why the RSDPE under a small measurement noise can not have a very accurate performance, which is merely close to the WLS-DRSS as shown in Fig. 3(b). However, the RSDPE surprisingly becomes the best estimator under a large measurement noise, as shown in Fig. 3(a). It seems that the RSDPE possesses a very good robustness to the deviation of the whitening procedure caused by the approximation inaccuracy in (5) under a large measurement noise. An interpretation for this is that this deviation yields the same impact as that of $\boldsymbol{\Delta}_{\boldsymbol{\Phi}}$. And the robustness to a bad anchor node placement is also shown in Fig. 2(c).

vi) *SDP-RSS:* The SDP-RSS yields the worst performance under a small measurement noise. Besides the fact that the SDP procedure yields a suboptimal solution, this is also because using the $\ell_1$-norm might not be the best choice for the SDP-RSS due to a lack of ML optimality. Under a large measurement noise, the SDP-RSS becomes better, almost the same as the A-BLUE. However, the high computational complexity brought by the SDP procedure and a lack of robustness make it lose its advantage over our proposed estimators.

*3) Impact of the Path-Loss Exponent:* We are also interested in how the PLE impacts our proposed estimators and hence we study our proposed estimators with a perfectly known data model under different PLEs when considering a large measurement noise. In fact, the PLE increases when the surrounding environment becomes more severe. Interestingly though, all the estimators grow more accurate in a more severe surrounding environment, as clearly depicted in Fig. 4. The performance of our proposed estimators can also be interpreted from our model error $\boldsymbol{v}$ in (7c), where the covariance of $\boldsymbol{v}$ obviously drops with an increasing PLE.

*4) Impact of Imperfect Path-Loss Exponent Estimate:* We previously assumed that the PLE $\gamma$ is perfectly known. However, in practice, the PLE is calibrated or estimated before the localization phase [27], [28]. Hence, we have to consider the case where the PLE is not perfectly known, i.e., the model uncertainty is considered. Therefore, to study the performance of our proposed estimators in such a case, we have also conducted two MC simulations, where for each trial an imperfect PLE $\widetilde{\gamma}$ is used to carry out the localization. The deviation $\Delta\gamma$, i.e., $\widetilde{\gamma} \triangleq \gamma + \Delta\gamma$, of the imperfect PLE from the actual PLE is considered to be zero-mean Gaussian distributed with variance $\sigma_\gamma^2$.

As shown in Fig. 5(a), all the estimators become worse with an increasing variance of the PLE estimate. The U-BLUE, the A-BLUE and the LE are all heavily impacted, while the RSDPE behaves relatively better, especially under a worse PLE estimate, due to its robust design.

To explain this in more detail, by using the imperfect PLE $\widetilde{\gamma}$, the imperfect $P'_{i,1}$ used to construct our data model in (4) is given by $\widetilde{P}'_{i,1} = 10^{\frac{P_{i,1}}{5(\gamma + \Delta\gamma)}}$. Using the first order Taylor series expansion of $\widetilde{P}'_{i,1}$ w.r.t. $\Delta\gamma$, we obtain

$$\widetilde{P}'_{i,1} = P'_{i,1}\left[1 - \frac{ln(10)P_{i,1}}{5\gamma^2}\Delta\gamma\right]. \tag{36}$$

Then, for a sufficiently small noise, (2) can be presented as $P_{i,1} \approx -5\gamma log_{10}\left(\frac{d_i^2}{d_1^2}\right)$ and hence (36) can be rewritten as

$$\widetilde{P}'_{i,1} \approx P'_{i,1}\left[1 + ln\left(\frac{d_i^2}{d_1^2}\right)\frac{\Delta\gamma}{\gamma}\right]. \tag{37}$$

Since $\Delta\gamma \sim \mathcal{N}(0, \sigma_\gamma^2)$, from (37), we can clearly see that an increasing variance $\sigma_\gamma^2$ of the imperfect PLE $\widetilde{\gamma}$ incurs a more severe impact on our proposed estimators. Fortunately, under a large PLE $\gamma$, the impact of $\sigma_\gamma^2$ becomes less severe than under a small PLE, which can also be seen from Fig. 5(b). Finally, to better serve the following discussions, we should emphasize again that, among all the estimators, the RSDPE yields the best performance in case of an imperfect PLE.

*5) Impact of Inaccurate Anchor Location Information:* In real life, the anchor location information might be inaccurate, if obtained by the global positioning system (GPS). Especially in military scenarios, this kind of information might be even more difficult to obtain, unreliable or tampered with by attackers. Therefore, we have to consider the model uncertainty in case of inaccurate anchor location information. Two MC simulations have been conducted, where for each trial every anchor location is given with a deviation, i.e., $\widetilde{\mathbf{s}}_i \triangleq \mathbf{s}_i + \boldsymbol{\delta}_s, \forall i$, where $\boldsymbol{\delta}_s \sim \mathcal{N}(\mathbf{0}, \sigma_s^2\mathbf{I}_d)$.

As shown in Fig. 5(c), all the estimators behave worse with an increasing variance of the anchor location inaccuracy, but the RSDPE again yields the best performance, due to its design for coping with model uncertainties. Finally, we notice that, if an inaccurate anchor location $\widetilde{\mathbf{s}}_i$ is used for constructing our data model in (4), considering the fact that $\boldsymbol{\delta}_s$ is scaled by $P'_{i,1}$, a large PLE will lead to a small value of $P'_{i,1}$ and hence can suppress the impact of $\boldsymbol{\delta}_s$, which can be observed in Fig. 5(d).

*F. Discussions*

In this subsection, we present the proposed estimators in a more general context. This discussion is also suitable for other localization problems, since there exist some common issues between the proposed localization problems and other ones.

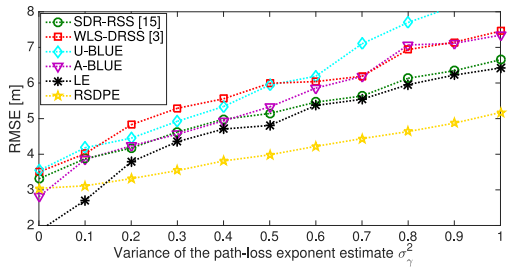

(a) Performance of different estimators under imperfect PLE knowledge when the actual PLE is $\gamma = 4$. The anchor location information is accurate.

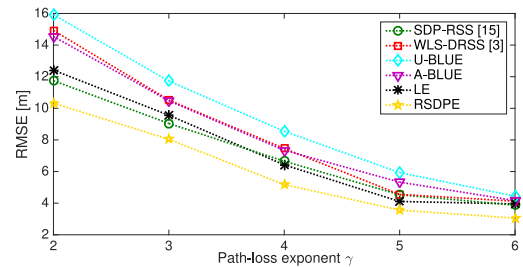

(b) Performance of different estimators with an imperfect PLE estimate under different PLEs when the variance of the PLE estimate is $\sigma_\gamma^2 = 1$. The anchor location information is accurate.

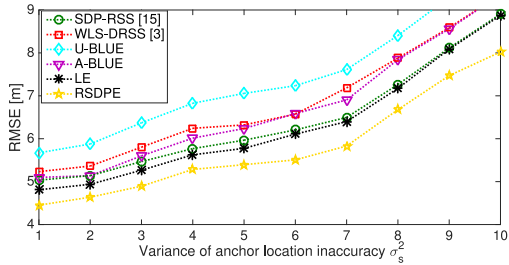

(c) Performance of different estimators with inaccurate anchor location information when the actual PLE is known and $\gamma = 4$.

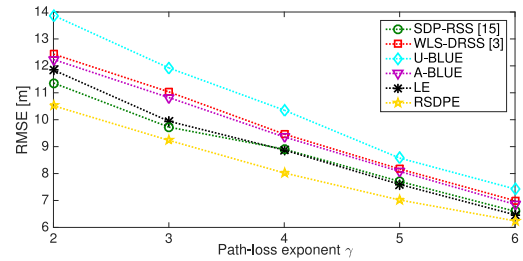

(d) Performance of different estimators under different PLEs when the actual PLE is known and the variance of the anchor location inaccuracy is $\sigma_s^2 = 10$.

Fig. 5. Performance comparison of different estimators with model uncertainties when the variance of the measurement noise is $\sigma_n^2 = 1$.

For optimal localization problems, non-linearity and non-convexity issues are inevitable, both of which are due to the distance norm $d_i = ||\mathbf{x} - \mathbf{s}_i||_2$. To be more specific, the distance norm is obviously non-linear w.r.t. $\mathbf{x}$, and the target cannot physically overlap with the anchors, i.e., $\mathbf{x} \neq \mathbf{s}_i, \forall i$, which explains the non-convexity. Most localization techniques first cope with non-linearity, either by directly applying a Taylor series expansion (TSE) around an appropriate value of $\mathbf{x}$, or by squaring and unfolding the distance norm. The former leads to some iterative ML methods, where a good initiation is critical for coping with the non-convexity as shown in Fig. 1(a). The latter one, which is our main focus, requires squaring the distance norm as $d_i^2 = ||\mathbf{x}||_2^2 - 2\mathbf{s}_i^T\mathbf{x} + ||\mathbf{s}_i||_2^2$, where $R \triangleq ||\mathbf{x}||_2^2$ has to be considered as a new unknown parameter to avoid non-linearity. As a result, a linear unconstrained localization problem can be formulated, which has $[\mathbf{x}^T, R]^T$ as a new unknown parameter vector (other unknown parameters could be jointly estimated as well), directly leading to a closed-form (weighted) LS solution. We categorize this kind of estimator as the unconstrained linear least squares estimator (ULLSE), and obviously the ULLSE ignores the fact that the new parameter vector $[\mathbf{x}^T, R]^T$ (or the one that contains it) is still bound to a non-convex set. To cope with that, the relation $R = \mathbf{x}^T\mathbf{x}$ should be considered, and accordingly other localization techniques can be considered:

1) The two-step linear least squares estimation (TLLSE) first obtains an initial estimate from the ULLSE and then fine-tunes it in the second step based on $R = \mathbf{x}^T\mathbf{x}$, equivalently the constraints in (10b) and (14). The Achilles' heel of the TLLSE are approximations like (12) or (15), which are often carried out to facilitate the update of the estimate. The goodness of such approximations often relies on the ULLSE. Under a small measurement noise and an exactly known data model, the ULLSE and hence the approximations are reliable, leading to a very good performance of the TLLSE. However, when the measurement noise becomes severe or there exist considerable model uncertainties, the approximations deteriorate, thus significantly undermining the performance of the TLLSE. In this paper, the A-BLUE tries to minimize the impact of the approximations as much as possible, e.g., by sticking to the original data model instead of constructing a new one. Please refer to Section III-A for details and references therein.

2) The semidefinite relaxation (SDR)-based estimator (SDRE) reformulates $R = \mathbf{x}^T\mathbf{x}$ as an LMI as in (26c) such that an SDP problem can be constructed. More importantly though, this also requires introducing slack variables so as to change the optimization problem from minimizing the cost function to its upper bound. All those procedures lead to a relatively worse performance of the SDRE under a small measurement noise, but guarantee a very good estimation accuracy under a large noise. In this paper, the RSDPE is particularly improved with a robust design against model uncertainties. Please refer to Section III-C for details and references therein.

3) The exact estimator (EE) is the theoretically optimal solution when considering the relation $R = \mathbf{x}^T\mathbf{x}$. The EE translates it into a new constraint as in (19b) without any approximation or dropping a constraint. Therefore, given an exactly known data model, if the solution that meets the KKT conditions can be precisely found, the EE should perform the best under both small and large measurement noises. It is worth noting that (19b) is still a non-convex constraint, which makes the search for the global solution very important. In this paper, the LE provides a useful interval, in which only the global solution resides. However, when the data model is uncertain, the global solution will be more difficult to find and hence the EE will not perform as good as expected. Please refer to Section III-B for details and references therein.

TABLE I
COMPARISON OF DIFFERENT KINDS OF ESTIMATION FOR LOCALIZATION PROBLEM

| | ULLSE | TLLSE | SDRE | EE |
|---|---|---|---|---|
| Examples[a] | U-BLUE, [15], [17][b], [18] | A-BLUE, [3], [16], [33], [34] | RSDPE, [15], [19], [20]–[23] | LE, [31] |
| Considering $R = \mathbf{x}^T\mathbf{x}$ | ✗ | ✓ | ✓ | ✓ |
| No approximation error[c] | - | ✗ | ✓ | ✓ |
| Considering all constraints | - | ✓ | ✗ | ✓ |
| Accuracy under a small measurement noise | Bad | Good | Medium | Very good |
| Accuracy under a large measurement noise | Very bad | Medium | Good | Good |
| Robustness to model uncertainties | Very bad | Bad | Good | Medium |
| Complexity | Very low | Low | High | Medium |

[a]For convenience, the examples are provided only in the field of RSS/DRSS-based localization, but the conclusions are not limited to this field.
[b][17] eliminates $R$ by taking differences between a selected reference and the other nodes, which is equivalent to ignoring $R = \mathbf{x}^T\mathbf{x}$.
[c]The approximation here only refers to the one related to the relation $R = \mathbf{x}^T\mathbf{x}$.

TABLE II
COMPARISON OF DIFFERENT METHODS FOR PLE CALIBRATION

| Methods \ Drawbacks | Anchor Dependence | Intensive Node Cooperation | Not Pervasive[a] |
|---|---|---|---|
| Anchor-Based [11], [51]–[53] | ✓ | ✓ | ✓ |
| Anchor-Free [26], [54] | ✗ | ✓ | ✓ |
| Self-Estimation [25] | ✗ | ✗ | ✓[b] |
| Collective Self-Estimation [27], [28] | ✗ | ✗ | ✗ |

[a]A pervasive method is a method that can be implemented in any kind of wireless network, i.e., without any external assistance or information.
[b]They still require some external information (e.g., network density) or a frequently changing receiver configuration and hence are not pervasive.

Localization techniques from the same category have a similar level of computational complexity and hence we can refer to Section III-D. We give a general comparison of the ULLSE, the TLLSE, the SDRE and the EE in Table I, where also some other examples beyond the proposed estimators are listed.

Also, it is very important to notice that most localization techniques (not limited to the RSS/DRSS-based) use a colored data model, which will generally degrade the localization performance and also explains why our proposed estimators are relatively better. Furthermore, some data models are very difficult or even impossible to whiten, since the true nodal distances might be required for whitening like the one in [3] and the famous Chan algorithm [17], [33]–[35]. Additionally, taking differences between the observations, e.g., generating TDOA or DRSS measurements, also leads to a colored model noise, which is often ignored in literature [17], [18], [33], [34], [50].

After all, it is hard to say which kind of estimator is overall the best. Based on Table I, we can choose the most suitable estimator or adaptively switch from one to another according to the demands. For example, if a low complexity is the most important consideration, the TLLSE could be the best choice. Under a severe measurement noise or given an uncertain data model, the SDRE is recommended. If there is no particular requirement, the EE is a good choice, since it has a good performance and yields the best accuracy under a small measurement noise.

IV. ESTIMATOR FOR UNKNOWN PATH-LOSS MODEL

In the previous sections, we have introduced robust DRSS-based localization for a known PLE. Based on these studies, we want to take one step further and explore a new iterative estimator which can jointly estimate the unknown PLE $\gamma$ and the unknown location $\mathbf{x}$.

A. Handling Unknown Path-Loss Exponent

Before introducing our new method, we would like to first discuss the current techniques to cope with an unknown PLE. Presently, many RSS/DRSS based localization methods assume a perfect pre-calibration stage without any PLE estimation error. Ironically though, PLE calibration techniques are still rarely studied. Here, we try to collect and summarize them in Table II. The anchor-based methods [11], [51]–[53] have to be carried out between the anchors and hence are very susceptible to inaccurate anchor location information. Based on some geometric constraints, the anchor-free methods [26], [54] can estimate the unknown PLE for temporarily grouped nodes without any location information. But, they still require intensive node cooperation and might cause a heavy network load. Therefore, if each node can self-estimate the PLE in a distributed fashion, this could solve the aforementioned issues [25]. Pervasiveness is another shortcoming which we have to conquer, since the PLE is a very crucial wireless channel parameter and we want to enable a collective PLE self-estimation [27], [28] that can be used in any kind of wireless device for facilitating efficient communication and networking designs. In a nutshell, a more robust and cost effective PLE pre-calibration stage can undoubtedly benefit the localization procedure.

Alternatively, we can conveniently skip the PLE pre-calibration when it is not available or reliable. Then, we have to jointly estimate the unknown PLE and the target location, which could also save a lot of resources. In this section, we are particularly interested in this kind of solution. Commonly, an initial guess for the unknown PLE $\gamma$ has to be adopted to obtain a quasi-estimate of the target location, which can then be used to update the PLE estimate [10], [17], [18], [22], [29]–[32]. Obviously, this will cause model uncertainties for the localization problem, which are often ignored however. Therefore, based on the previous studies, we want to seek a new robust DRSS localization approach in case of an unknown PLE.

### B. Prototype of the Proposed Iterative Estimator

In addition to the model in (4), if given a known target location, we can obtain another linear model from (2) as

$$\boldsymbol{\pi} = \boldsymbol{\lambda}\gamma + \boldsymbol{\nu}, \tag{38}$$

where $\boldsymbol{\pi} \triangleq [\cdots, P_{i,1}, \cdots]^T$, $\boldsymbol{\nu} \triangleq [\cdots, n_{i,1}, \cdots]^T$ and $\boldsymbol{\lambda} \triangleq [\cdots, -10\log_{10}\left(\frac{\|\mathbf{x}-\mathbf{s}_i\|_2}{\|\mathbf{x}-\mathbf{s}_1\|_2}\right), \cdots]^T$. Again, we stack the equations for a fixed RN and all anchor nodes $i \neq 1$. However, it is very difficult to obtain a single linear model for both an unknown target location and an unknown PLE. This enlightens us that a block coordinate descent (BCD) method might be applicable to this problem [55]. In order to do so, we need to redefine the parameter vector $\boldsymbol{\theta}$ to be estimated as $\boldsymbol{\theta} \triangleq [\mathbf{x}^T, \|\mathbf{x}\|_2^2, \gamma]^T$. The BCD is implemented by partitioning $\boldsymbol{\theta}$ into two blocks, $[\mathbf{x}^T \ \|\mathbf{x}\|_2^2]^T$ and $\gamma$, and then at each iteration a cost function is minimized with respect to one of the blocks while the other is held fixed. We denote the $\boldsymbol{\theta}$ estimate at the $k$-th iteration as $\hat{\boldsymbol{\theta}}^{(k)}$, the iteration tolerance as $\xi$ and the cost functions for estimating the target location and the PLE respectively as $J'(\cdot)$ and $J''(\cdot)$. The prototype of our proposed estimator is presented in Algorithm 1.

### C. Robust Semidefinite Programming Based Block Coordinate Descent Estimator

To fully describe our method, we need to elaborate on the minimizations in (39) and (40). Since the RSDPE has a very good robustness to imperfect PLE knowledge, applying a similar method to (38) might also result in a good robustness to imperfect target location knowledge. Therefore, the idea behind our robust SDP-based block coordinate descent estimator (RSDP-BCDE) is to utilize this method to update both the location and the PLE. Considering that our method introduces two new auxiliary variables, next to the parameter vector $\boldsymbol{\theta}$, we introduce the slack variables $t_1$, $\alpha_1$ and $t_2$, $\alpha_2$ to update the target location estimate and the PLE estimate, respectively. Additionally, two bounds $\zeta_1$ and $\zeta_2$ are also needed, which are both computed in the same way as the RSDPE does for $\zeta$.

For updating the block $[\mathbf{x}^T, \|\mathbf{x}\|_2^2]^T$, we use

$$\left([\hat{\boldsymbol{\theta}}^{(k+1)}]_{1:d+1}, \hat{t}_1^{(k+1)}, \hat{\alpha}_1^{(k+1)}\right) = \arg\min_{[\boldsymbol{\theta}]_{1:d+1}, t_1, \alpha_1} t_1$$

---

**Algorithm 1:** PROTOTYPE of proposed iterative estimator.

1 *Initialization:* Choose the initial value $\hat{\boldsymbol{\theta}}^{(0)}$;
2 *Loop:* Given $\hat{\boldsymbol{\theta}}^{(k)} = [[\hat{\boldsymbol{\theta}}^{(k)}]_{1:d+1}^T, [\hat{\boldsymbol{\theta}}^{(k)}]_{d+2}]^T$, solve

$$\left[\hat{\boldsymbol{\theta}}^{(k+1)}\right]_{1:d+1} = \arg\min_{[\boldsymbol{\theta}]_{1:d+1}} J'\left([\boldsymbol{\theta}]_{1:d+1}, [\hat{\boldsymbol{\theta}}^{(k)}]_{d+2}\right); \tag{39}$$

3 Given $[[\hat{\boldsymbol{\theta}}^{(k+1)}]_{1:d+1}^T, [\hat{\boldsymbol{\theta}}^{(k)}]_{d+2}]^T$, solve

$$\left[\hat{\boldsymbol{\theta}}^{(k+1)}\right]_{d+2} = \arg\min_{[\boldsymbol{\theta}]_{d+2}} J''\left([\hat{\boldsymbol{\theta}}^{(k+1)}]_{1:d+1}, [\boldsymbol{\theta}]_{d+2}\right); \tag{40}$$

4 Let $\hat{\boldsymbol{\theta}}^{(k+1)} = [[\hat{\boldsymbol{\theta}}^{(k+1)}]_{1:d+1}^T, [\hat{\boldsymbol{\theta}}^{(k+1)}]_{d+2}]^T$;
5 If $\|[\hat{\boldsymbol{\theta}}^{(k+1)}]_{1:d} - [\hat{\boldsymbol{\theta}}^{(k)}]_{1:d}\|_2 \leq \xi$, continue. Otherwise go back to *Loop*;
6 **return** $\hat{\boldsymbol{\theta}}^{(k+1)}$;

---

subject to

$$\begin{bmatrix} (1-\alpha_1)\mathbf{I}_{N-1} & \widetilde{\boldsymbol{\Phi}}^{(k)}[\boldsymbol{\theta}]_{1:d+1} - \widetilde{\boldsymbol{\rho}}^{(k)} & \mathbf{0} \\ (\widetilde{\boldsymbol{\Phi}}^{(k)}[\boldsymbol{\theta}]_{1:d+1} - \widetilde{\boldsymbol{\rho}}^{(k)})^T & t_1 & -\zeta_1[\boldsymbol{\theta}]_{1:d+1}^T \\ \mathbf{0} & -\zeta_1[\boldsymbol{\theta}]_{1:d+1} & \alpha_1\mathbf{I}_{d+1} \end{bmatrix}$$
$$\succeq \mathbf{0},$$

$$\begin{bmatrix} \mathbf{I}_{d\times d} & [\boldsymbol{\theta}]_{1:d} \\ [\boldsymbol{\theta}]_{1:d}^T & [\boldsymbol{\theta}]_{d+1} \end{bmatrix} \succeq \mathbf{0}, \tag{41}$$

where $\hat{t}_1^{(k+1)}$ and $\hat{\alpha}_1^{(k+1)}$ are respectively the estimates of $t_1$ and $\alpha_1$ at the $(k+1)$-th iteration, $\widetilde{\boldsymbol{\Phi}}^{(k)}$ and $\widetilde{\boldsymbol{\rho}}^{(k)}$ are respectively the $\boldsymbol{\Phi}$ and the $\boldsymbol{\rho}$ constructed by the imperfect PLE estimate at the $k$-th iteration, i.e., $[\hat{\boldsymbol{\theta}}^{(k)}]_{d+2}$.

For updating $\gamma$, we notice from (38) that $\boldsymbol{\nu} = \boldsymbol{\Gamma}\mathbf{n}$ and hence the covariance matrix of $\boldsymbol{\nu}$ is $\boldsymbol{\Sigma}_{\boldsymbol{\nu}} = \sigma_n^2 \boldsymbol{\Gamma}\boldsymbol{\Gamma}^T$. Hence, the whitened model of (38) can be expressed as

$$\boldsymbol{\Sigma}_{\boldsymbol{\nu}}^{-1/2}\boldsymbol{\pi} = \boldsymbol{\Sigma}_{\boldsymbol{\nu}}^{-1/2}\boldsymbol{\lambda}\gamma + \boldsymbol{\Sigma}_{\boldsymbol{\nu}}^{-1/2}\boldsymbol{\nu} \tag{42a}$$

$$\Rightarrow (\boldsymbol{\Gamma}\boldsymbol{\Gamma}^T)^{-1/2}\boldsymbol{\pi} = (\boldsymbol{\Gamma}\boldsymbol{\Gamma}^T)^{-1/2}\boldsymbol{\lambda}\gamma + (\boldsymbol{\Gamma}\boldsymbol{\Gamma}^T)^{-1/2}\boldsymbol{\nu} \tag{42b}$$

$$\Rightarrow \mathbf{c} = \mathbf{d}\gamma + \mathbf{e}, \tag{42c}$$

where $\mathbf{c} \triangleq (\boldsymbol{\Gamma}\boldsymbol{\Gamma}^T)^{-1/2}\boldsymbol{\pi}$, $\mathbf{d} \triangleq (\boldsymbol{\Gamma}\boldsymbol{\Gamma}^T)^{-1/2}\boldsymbol{\lambda}$ and $\mathbf{e} \triangleq (\boldsymbol{\Gamma}\boldsymbol{\Gamma}^T)^{-1/2}\boldsymbol{\nu}$. Note that now the covariance matrix of $\mathbf{e}$ is $\boldsymbol{\Sigma}_\mathbf{e} = \sigma_n^2 \mathbf{I}_N$. Based on the whitened data model (42c), we update $\gamma$ as

$$\left([\hat{\boldsymbol{\theta}}^{(k+1)}]_{d+2}, \hat{t}_2^{(k+1)}, \hat{\alpha}_2^{(k+1)}\right) = \arg\min_{[\boldsymbol{\theta}]_{d+2}, t_2, \alpha_2} t_2$$

subject to

$$\begin{bmatrix} (1-\alpha_2)\mathbf{I}_{N-1} & \mathbf{c} - \tilde{\mathbf{d}}^{(k+1)}[\boldsymbol{\theta}]_{d+2} & \mathbf{0} \\ (\mathbf{c} - \tilde{\mathbf{d}}^{(k+1)}[\boldsymbol{\theta}]_{d+2})^T & t_2 & -\zeta_2[\boldsymbol{\theta}]_{d+2}^T \\ \mathbf{0} & -\zeta_2[\boldsymbol{\theta}]_{d+2} & \alpha_2\mathbf{I}_{d+1} \end{bmatrix} \succeq \mathbf{0},$$
$$\tag{43}$$

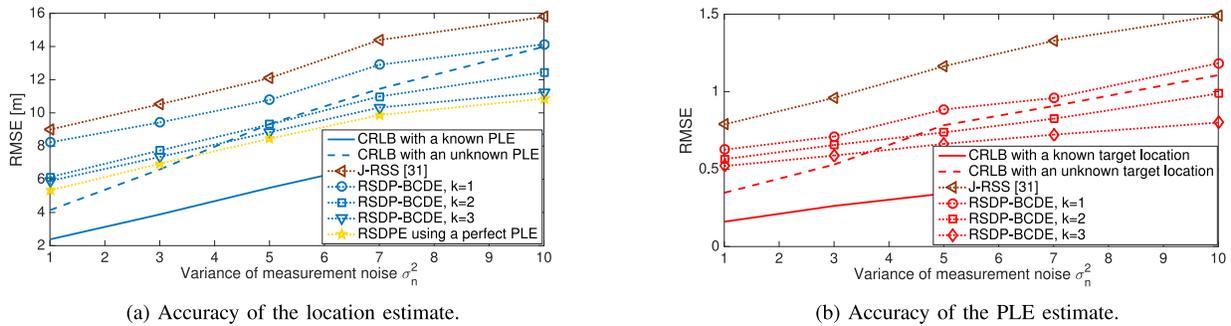

(a) Accuracy of the location estimate.

(b) Accuracy of the PLE estimate.

Fig. 6. Performance of our proposed RSDP-BCDE under different noise conditions: the PLE is $\gamma = 2$; the initial value of the PLE estimate is $\hat{\gamma}^{(0)} = 4$; $k$ is the iteration number when the RSDP-BCDE stops the iterative procedure.

where $\hat{t}_2^{(k+1)}$ and $\hat{\alpha}_2^{(k+1)}$ are respectively the estimates of $t_2$ and $\alpha_2$ at the $(k+1)$-th iteration, $\tilde{\mathbf{d}}^{(k+1)}$ is the $\mathbf{d}$ constructed by the imperfect target location estimate at the $(k+1)$-th iteration, i.e., $[\hat{\boldsymbol{\theta}}^{(k+1)}]_{1:d}$.

Finally, the optimization problems (41) and (43) are solved by *CVX* and the complexity of the RSDP-BCDE after $k$ iterations is $O[kd^2 N^{2.5} ln(1/\xi)]$.

### D. Numerical Results

To study the performance of the RSDP-BCDE, we have conducted an MC simulation. We select the initial value of the PLE estimate as 4 considering that the PLE normally ranges from 2 to 6 [6]. In the simulation, the PLE is set to 2 and the rest of the MC simulation settings are the same as before. The numerical results are shown in Fig. 6.

The RSDP-BCDE is studied for different iteration numbers $k$ and compared against one of the RSS-based estimators (J-RSS) from [31], which jointly estimates the unknown transmit power and PLE. According to the simulation results, our proposed method outperforms the J-RSS. As shown in Fig. 6(a), with more iterations, the performance of the RSDP-BCDE becomes better and gradually approaches that of the RSDPE using a perfect $\gamma$. The PLE estimate also becomes more accurate with an increasing iteration number $k$, as shown in Fig. 6(b). We also notice that, after the first iteration, the performance of the RSDP-BCDE is already very close to the CRLB with an unknown PLE or target location due to a good initial value of the PLE. Then, with more iterations, the knowledge of the target location and the PLE becomes better, thus improving the performance of the RSDP-BCDE over the CRLB with unknown PLE or target location. Additionally, the RSDP-BCDE converges quickly under a small measurement noise.

To end this section, we can conclude from the numerical results that even if the path-loss model is unknown, the RSDP-BCDE is still able to obtain an accurate location estimate. However, note that the SDP procedure has a very large complexity in each iteration. Hence, if the PLE is already accurate enough, we can similarly replace the SDP procedure with the A-BLUE or the LE to estimate the location such that the total computational complexity can be greatly reduced.

## V. CONCLUSION

A whitened model for DRSS-based localization has been introduced and studied. Based on such a model, we have proposed and analyzed three different estimators for a known path-loss model (i.e., the A-BLUE, the LE and the RSDPE), where the latter is robust against an imperfect PLE estimate or inaccurate anchor location information. We have also proposed one robust iterative estimator for an unknown path-loss model (i.e., the RSDP-BCDE).

Simulation results have shown that, when the PLE is known, our three proposed estimators outperform an RSS-based joint estimator (SDP-RSS), which applies the SDP-procedure on an $\ell_1$-norm approximation, as well as a recent weighted least squares estimator (WLS-DRSS), which requires perfect knowledge of the variance of the measurement noise. The performance of our three proposed estimators for a known PLE is studied under different noise conditions, different PLEs, imperfect PLE knowledge and inaccurate anchor location information. Their computational complexities are also investigated. Each estimator has its own advantages: the A-BLUE has the lowest computational complexity; the LE yields the best performance for a small measurement noise; and the RSDPE holds the best accuracy under a large measurement noise, an imperfect PLE and inaccurate anchor location information. Besides, in case of an unknown PLE, it is finally shown that, with more iterations, the performance of the RSDP-BCDE can approach that of the RSDPE with a known path-loss model. In real-life, to meet different practical demands when encountering different situations, different proposed estimators are provided as options.

## APPENDIX A
## RSS COLLECTION

Assume the received signal $y(t)$ with the time index $t$ can be expressed as

$$y(t) = x(t) \star h(t) + n(t), \quad (A.1)$$

where $\star$ denotes the convolution operator, $x(t)$ is the transmitted signal, $h(t)$ indicates the channel response and $n(t)$ is the additive zero-mean white *Gaussian* noise. In most literature, the RSS refers to the signal power after a successful demodulation. To be specific, if the signal $y(t)$ can successfully be demodulated, $n(t)$ is cancelled and hence we can easily observe the signal envelope $r(t) \triangleq |x(t) \star h(t)|$, from which the RSS can be computed. In [4], [5], real-life experiments have been conducted to collect RSS measurements from demodulated signals, thereby demonstrating that the noise can be ignored. Moreover, the case of no signal demodulation is also investigated therein, but we feel this is beyond the scope of this paper.

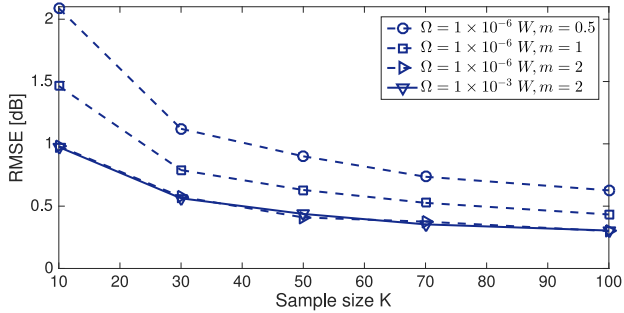

Fig. 7. Performance of RSS collection under different sample sizes and different values of $m$ and $\Omega$.

To further explain the RSS collection procedure, we notice that $r(t)$ is affected by *small-scale* fading. For simplicity, we will remove the time index $t$ from now on to represent an instantaneous value. If the *Nakagami-m* distribution is considered, which actually characterizes the instantaneous signal envelope $r$, the instantaneous received power $p \triangleq r^2$ is *Gamma* distributed as

$$\mathbb{P}(p|\Omega) = \frac{1}{\Gamma(m)} \left(\frac{m}{\Omega}\right)^m p^{m-1} e^{-\frac{mp}{\Omega}}. \qquad (A.2)$$

where $m$ is the fading parameter and a small value of $m$ indicates a severe fading. The other parameter $\Omega$, defined as $\Omega \triangleq E(p)$, is the RSS to collect, but expressed in *watts*. Therefore, collecting RSS measurements corresponds to estimating $\Omega$.

Denoting $\Omega_i$ as the $\Omega$ associated with the $i$-th anchor, the maximum likelihood (ML) estimate of $\Omega_i$ is readily given by

$$\widehat{\Omega}_i = \frac{1}{K} \sum_{k=1}^{K} p_i^{(k)}, \qquad (A.3)$$

where $p_i^{(k)}, \forall k = 1, \ldots, K$, represent $K$ consecutive samples of $p$ related to the $i$-th anchor. Due to the fact that $Var(p) = \frac{\Omega^2}{m}$, we can easily obtain

$$Var(\widehat{\Omega}_i) = \frac{\Omega_i^2}{Km}, \qquad (A.4)$$

which indicates that the estimation error can be reduced by taking more samples. Obviously, the ML estimate $\widehat{\Omega}_i$ is unbiased and, denoting $\widehat{\Omega}_i = \Omega_i + \Delta\Omega_i$, $\Delta\Omega_i$ is asymptotically *Normal* distributed as $\Delta\Omega_i \sim \mathcal{N}(0, \sigma^2_{\Delta\Omega_i})$ with $\sigma^2_{\Delta\Omega_i} = \frac{\Omega_i^2}{Km}$.

Expressing the RSS estimate in *dB* and applying the first-order Taylor series expansion w.r.t. $\Delta\Omega_i$ results in

$$\widehat{P}_i = 10 log_{10}(\widehat{\Omega}_i) = 10 log_{10}(\Omega_i + \Delta\Omega_i) \approx P_i + \Delta P_i, \quad (A.5)$$

where the estimation error of $\widehat{P}_i$ is denoted as $\Delta P_i \triangleq \frac{10}{ln(10)} \frac{\Delta\Omega_i}{\Omega_i}$ and hence $\Delta P_i \sim \mathcal{N}(0, \sigma^2_{\Delta P_i})$ with $\sigma^2_{\Delta P_i} = \frac{100}{ln(10)^2 Km}$. We notice that, compared with $\sigma^2_{\Delta\Omega_i}$, $\sigma^2_{\Delta P_i}$ does not depend on $\Omega_i$ any more, i.e., the RSS estimation error in dB is independent of the anchors. This means that, even if not enough samples are collected, the impact of $\Delta P_i$ is still similar to that of the shadowing effect $\chi_i$. We have also conducted a simple simulation for RSS collection. As shown in Fig. 7, the collection error decreases with a large $m$ and more samples. But more importantly, different values of $\Omega$ yield no significant impact on the collection error if considered in *dB*. In a nutshell, we can assume the RSS is perfectly collected in this paper without loss of generality.

## APPENDIX B
### DERIVATION FROM RSS-BASED MODEL

In this appendix, we show that our whitened DRSS-based model can also be derived from a properly whitened RSS-based model after orthogonally projecting out the unknown $\bar{P}'_0$.

To show that, let us first rewrite (1) as

$$||\mathbf{x} - \mathbf{s}_i||_2^2 = \frac{\bar{P}'_0 \Delta P'_{0,i} \chi'_i}{P'_i}, \qquad (B.1)$$

where $P'_i \triangleq 10^{\frac{P_i}{5\gamma}}$, $\bar{P}'_0 \triangleq 10^{\frac{\bar{P}_0}{5\gamma}}$, $\Delta P'_{0,i} \triangleq 10^{\frac{\Delta P_{0,i}}{5\gamma}}$, $\chi'_i \triangleq 10^{\frac{\chi_i}{5\gamma}}$ and the reference distance is again $d_0 = 1\ m$ without loss of generality. For a sufficiently small noise and using the first-order Taylor series expansion on (B.1), we obtain

$$||\mathbf{x}||_2^2 - 2\mathbf{s}_i^T \mathbf{x} + ||\mathbf{s}_i||_2^2 = \frac{\bar{P}'_0}{P'_i}\left[1 + \frac{ln(10)}{5\gamma} n_i\right], \quad (B.2)$$

where $n_i = \Delta P_{0,i} + \chi_i$. Then, we can formulate a linear model as

$$\mathbf{B}\boldsymbol{\phi} = \mathbf{h} + \boldsymbol{\varsigma} \qquad (B.3)$$

where

$$\mathbf{B} \triangleq \begin{bmatrix} \vdots & \vdots & \vdots \\ 2\mathbf{s}_i^T & -1 & 1/P'_i \\ \vdots & \vdots & \vdots \end{bmatrix},$$

$\boldsymbol{\phi} \triangleq [\mathbf{x}, ||\mathbf{x}||_2^2, \bar{P}'_0]^T$, $\mathbf{h} \triangleq [\cdots, ||\mathbf{s}_i||_2^2, \cdots]^T$ and $\boldsymbol{\varsigma} \triangleq [\cdots, -\frac{ln(10)\bar{P}'_0}{5\gamma P'_i} n_i, \cdots]^T$. Every element of $\boldsymbol{\varsigma}$, say $\varsigma_i$, is a zero-mean Gaussian variable with variance $\frac{[ln(10)]^2 \bar{P}'^2_0 \sigma^2_n}{25\gamma^2 P'^2_i}$ and hence the covariance matrix of $\boldsymbol{\varsigma}$ can be expressed as $\boldsymbol{\Sigma}_\varsigma = \frac{ln(10)^2 \bar{P}'^2_0 \sigma^2_n}{25\gamma^2} \mathbf{D}'^{-2}$, where $\mathbf{D}' = \text{diag}([P'_1, \cdots, P'_N]^T)$ with $\text{diag}(\cdot)$ a diagonal matrix with its argument on the diagonal.

Let us first describe the relation between the model in (4) and the RSS-based model in (B.3). By recalling the definitions of $\boldsymbol{p}$ and $\mathbf{h}$, we can easily observe that $\boldsymbol{p} = \mathbf{P}\mathbf{h}$, where $\mathbf{P} \triangleq \begin{bmatrix} \mathbf{1}_{(N-1)\times 1} & \text{diag}([-P'_{2,1}, \cdots, -P'_{N,1}]^T) \end{bmatrix} = -\frac{1}{P'_1}\boldsymbol{\Gamma}\mathbf{D}'$, and similarly, $\mathbf{P}\mathbf{B}\boldsymbol{\phi} = \boldsymbol{\Psi}\boldsymbol{\theta}$ and $\mathbf{P}\boldsymbol{\varsigma} = \boldsymbol{\epsilon}$.

Hence, before the whitening procedure, the DRSS-based model (4) can be viewed as the RSS-based model (B.3), where we remove the influence of $\bar{P}'_0$ by applying the transformation matrix $\mathbf{P}$. However, it is hard to judge at this point whether this operation will cause a loss of information or not.

In order to do that, let us first whiten the RSS-based model (B.3), which leads to

$$\boldsymbol{\Sigma}_{\boldsymbol{\varsigma}}^{-1/2} \mathbf{B} \boldsymbol{\phi} = \boldsymbol{\Sigma}_{\boldsymbol{\varsigma}}^{-1/2} \mathbf{h} + \boldsymbol{\Sigma}_{\boldsymbol{\varsigma}}^{-1/2} \boldsymbol{\varsigma} \quad \text{(B.4a)}$$

$$\Rightarrow \mathbf{D}' \mathbf{B} \boldsymbol{\phi} = \mathbf{D}' \mathbf{h} + \mathbf{D}' \boldsymbol{\varsigma} \quad \text{(B.4b)}$$

$$\Rightarrow \mathbf{B}' \boldsymbol{\phi} = \mathbf{h}' + \boldsymbol{\varsigma}', \quad \text{(B.4c)}$$

where $\mathbf{B}' \triangleq \mathbf{D}'\mathbf{B}$, $\mathbf{h}' \triangleq \mathbf{D}'\mathbf{h}$ and $\boldsymbol{\varsigma}' \triangleq \mathbf{D}'\boldsymbol{\varsigma}$ with the covariance matrix $\boldsymbol{\Sigma}_{\boldsymbol{\varsigma}'} = \frac{\ln(10)^2 \bar{P}_0'^2 \sigma_n^2}{25 \gamma^2} \mathbf{I}_N$.

Now, to see the relation between our whitened DRSS-based model (7) and this whitened RSS-based model ((B.4)), we can show that $\boldsymbol{\Phi}\boldsymbol{\theta} = (\boldsymbol{\Gamma}\boldsymbol{\Gamma}^T)^{-1/2} \mathbf{P} \mathbf{B} \boldsymbol{\phi} = \mathbf{P}' \mathbf{B}' \boldsymbol{\phi}$, where $\mathbf{P}' \triangleq -\frac{1}{P_1'}(\boldsymbol{\Gamma}\boldsymbol{\Gamma}^T)^{-1/2}\boldsymbol{\Gamma}$. So, after the whitening procedure, the whitened DRSS-based model (7) can be viewed as the whitened RSS-based model ((B.4)), where we remove the influence of $\bar{P}_0'$ by applying the transformation matrix $\mathbf{P}'$. The crucial observation now is that this transformation matrix $\mathbf{P}'$ is a (scaled) unitary operator, i.e., $\mathbf{P}'\mathbf{P}'^T = \frac{1}{P_1'^2}\mathbf{I}_{N \times N}$, and hence by taking differences of RSSs to eliminate the unknown transmit power, our whitened DRSS-based model does not entail any loss of information compared to the whitened RSS-based model.

## APPENDIX C
## CRAMÉR-RAO LOWER BOUNDS

To derive the Cramér-Rao lower bounds (CRLBs) used in this paper, we recall from (2) that the vector of DRSS samples, say $\boldsymbol{\pi}$, is Gaussian distributed as $\boldsymbol{\pi} \sim \mathcal{N}(\boldsymbol{\mu}, \boldsymbol{\Sigma}_{\boldsymbol{\pi}})$, where for $i \neq 1$ we have $\boldsymbol{\pi} \triangleq [\cdots, P_{i,1}, \cdots]^T$, $\boldsymbol{\mu} = [\cdots, \mu_i, \cdots]^T$ with $\mu_i = -10\gamma \log_{10}\left(\frac{||\mathbf{x}-\mathbf{s}_i||_2}{||\mathbf{x}-\mathbf{s}_1||_2}\right)$ and according to (6), $\boldsymbol{\Sigma}_{\boldsymbol{\pi}} = \sigma_n^2 \boldsymbol{\Gamma}\boldsymbol{\Gamma}^T$.

To obtain the CRLB, the Fisher information matrix (FIM) can be computed as [56]

$$[\mathbf{J}]_{n,m} = \left[\frac{\partial \boldsymbol{\mu}}{\partial \theta_n}\right]^T \boldsymbol{\Sigma}_{\boldsymbol{\pi}}^{-1} \left[\frac{\partial \boldsymbol{\mu}}{\partial \theta_m}\right] + \frac{1}{2}tr[\boldsymbol{\Sigma}_{\boldsymbol{\pi}}^{-1}\frac{\partial \boldsymbol{\Sigma}_{\boldsymbol{\pi}}}{\partial \theta_n}\boldsymbol{\Sigma}_{\boldsymbol{\pi}}^{-1}\frac{\partial \boldsymbol{\Sigma}_{\boldsymbol{\pi}}}{\partial \theta_m}], \quad \text{(C.1)}$$

where depending on the scenarios $\boldsymbol{\theta} = \mathbf{x}$, $\boldsymbol{\theta} = [\mathbf{x}^T, \gamma]^T$, or $\boldsymbol{\theta}$ is the scaler $\boldsymbol{\theta} = \gamma$, and $\frac{\partial \boldsymbol{\mu}}{\partial \theta_n} \triangleq [\cdots, \frac{\partial [\boldsymbol{\mu}]_i}{\partial \theta_n}, \cdots]^T$. Since $\boldsymbol{\Sigma}_{\boldsymbol{\pi}}$ does not depend on $\boldsymbol{\theta}$, we can simplify (C.1) as $[\mathbf{J}]_{n,m} = \left[\frac{\partial \boldsymbol{\mu}}{\partial \theta_n}\right]^T \boldsymbol{\Sigma}_{\boldsymbol{\pi}}^{-1} \left[\frac{\partial \boldsymbol{\mu}}{\partial \theta_m}\right]$. Letting $\mathbf{x} = [x_1, \cdots, x_d]^T$ and $\mathbf{s}_i = [s_{i,1}, \cdots, s_{i,d}]^T$, we obtain

$$\frac{\partial [\boldsymbol{\mu}]_i}{\partial x_k} = -\frac{10\gamma}{\ln(10)} \\ \times \frac{(x_k - s_{i,k})||\mathbf{x}-\mathbf{s}_1||_2^2 - (x_k - s_{1,k})||\mathbf{x}-\mathbf{s}_i||_2^2}{||\mathbf{x}-\mathbf{s}_i||_2^2 ||\mathbf{x}-\mathbf{s}_1||_2^2}, \\ k = 1, \cdots, d \quad \text{(C.2)}$$

and $\frac{\partial [\boldsymbol{\mu}]_i}{\partial \gamma} = -10\log_{10}\left(\frac{||\mathbf{x}-\mathbf{s}_i||_2}{||\mathbf{x}-\mathbf{s}_1||_2}\right)$.

*a) CRLBs on Joint Location Estimate and PLE Estimate:* In this case, $\boldsymbol{\theta} = [\mathbf{x}^T, \gamma]^T$ in $\mathbb{R}^{d+1}$ and the CRLB for the location

estimate is obtained as

$$CRLB_1 = \sqrt{\sum_{k=1}^{d}[\mathbf{J}^{-1}]_{k,k}},$$

while the CRLB for the PLE estimate is obtained as

$$CRLB_2 = \sqrt{[\mathbf{J}^{-1}]_{d+1,d+1}}.$$

*b) CRLB on Location Estimate with a Known PLE:* In this case, $\boldsymbol{\theta} = \mathbf{x}$ in $\mathbb{R}^d$ and the CRLB for the location estimate with a known PLE is obtained as

$$CRLB_3 = \sqrt{\sum_{k=1}^{d}[\mathbf{J}^{-1}]_{k,k}}.$$

*c) CRLB on PLE estimate with a known location:* In this case, $\boldsymbol{\theta} = \gamma$ and the CRLB for the PLE estimate with a known location is simply given by

$$CRLB_4 = \sqrt{1/[\mathbf{J}]_{1,1}},$$

where we note that $\mathbf{J}$ is just a scaler here.

## ACKNOWLEDGMENT

The authors would like to sincerely thank the anonymous reviewers for their valuable advice.

## REFERENCES

[1] N. Patwari, J. Ash, S. Kyperountas, A. Hero, R. Moses, and N. Correal, "Locating the nodes: Cooperative localization in wireless sensor networks," *IEEE Signal Process. Mag.*, vol. 22, no. 4, pp. 54–69, Jul. 2005.

[2] K. Whitehouse, C. Karlof, and D. Culler, "A practical evaluation of radio signal strength for ranging-based localization," *SIGMOBILE Mob. Comput. Commun. Rev.*, vol. 11, no. 1, pp. 41–52, Jan. 2007.

[3] L. Lin, H. So, and Y. Chan, "Accurate and simple source localization using differential received signal strength," *Digit. Signal Process.*, vol. 23, no. 3, pp. 736–743, 2013.

[4] R. Martin, A. King, J. Pennington, R. Thomas, R. Lenahan, and C. Lawyer, "Modeling and mitigating noise and nuisance parameters in received signal strength positioning," *IEEE Trans. Signal Process.*, vol. 60, no. 10, pp. 5451–5463, Oct. 2012.

[5] R. K. Martin, R. W. Thomas, and Z. Wu, "Using spectral correlation for non-cooperative RSS-based positioning," in *Proc. IEEE Stat. Signal Process. Workshop*, Jun. 2011, pp. 241–244.

[6] T. Rappaport, *Wireless Communications: Principles and Practice*, 2nd ed. Upper Saddle River, NJ, USA: Prentice-Hall, 2001.

[7] B. Sklar, "Rayleigh fading channels in mobile digital communication systems. I. Characterization," *IEEE Commun. Mag.*, vol. 35, no. 7, pp. 90–100, Jul. 1997.

[8] N. Nakagami, "The m-distribution, a general formula for intensity distribution of rapid fading," *Stat. Methods Radio Wave Propag.*, W. G. Hoffman, Ed. Oxford, U.K.: Pergamon, 1960.

[9] S. Boyd and L. Vandenberghe, *Convex Optimization*. Cambridge, U.K.: Cambridge Univ. Press, 2004.

[10] X. Li, "RSS-based location estimation with unknown pathloss model," *IEEE Trans. Wireless Commun.*, vol. 5, no. 12, pp. 3626–3633, Dec. 2006.

[11] A. Coluccia and F. Ricciato, "On ML estimation for automatic RSS-based indoor localization," in *Proc. 5th IEEE Int. Symp. Wireless Pervasive Comput.*, May 2010, pp. 495–502.

[12] N. Patwari, A. Hero, M. Perkins, N. Correal, and R. O'Dea, "Relative location estimation in wireless sensor networks," *IEEE Trans. Signal Process.*, vol. 51, no. 8, pp. 2137–2148, Aug. 2003.

[13] X. Li, "Collaborative localization with received-signal strength in wireless sensor networks," *IEEE Trans. Veh. Technol.*, vol. 56, no. 6, pp. 3807–3817, Nov. 2007.

[14] J. H. Lee and R. M. Buehrer, "Location estimation using differential RSS with spatially correlated shadowing," in *Proc. IEEE Global Telecommun. Conf.*, Nov. 2009, pp. 1–6.

[15] R. Vaghefi, M. Gholami, and E. Strom, "RSS-based sensor localization with unknown transmit power," in *Proc. IEEE Int. Conf. Acoust., Speech, Signal Process.*, May 2011, pp. 2480–2483.

[16] H. C. So and L. Lin, "Linear least squares approach for accurate received signal strength based source localization," *IEEE Trans. Signal Process.*, vol. 59, no. 8, pp. 4035–4040, Aug. 2011.

[17] N. Salman, A. Kemp, and M. Ghogho, "Low complexity joint estimation of location and path-loss exponent," *IEEE Wireless Commun. Lett.*, vol. 1, no. 4, pp. 364–367, Aug. 2012.

[18] Y. Xu, J. Zhou, and P. Zhang, "RSS-based source localization when path-loss model parameters are unknown," *IEEE Commun. Lett.*, vol. 18, no. 6, pp. 1055–1058, Jun. 2014.

[19] C. Meng, Z. Ding, and S. Dasgupta, "A semidefinite programming approach to source localization in wireless sensor networks," in *IEEE Signal Process. Lett.*, vol. 15, pp. 253–256, 2008, doi: 10.1109/LSP.2008.916731.

[20] R. Ouyang, A.-S. Wong, and C.-T. Lea, "Received signal strength-based wireless localization via semidefinite programming: Noncooperative and cooperative schemes," *IEEE Trans. Veh. Technol.*, vol. 59, no. 3, pp. 1307–1318, Mar. 2010.

[21] R. Vaghefi, M. Gholami, R. Buehrer, and E. Strom, "Cooperative received signal strength-based sensor localization with unknown transmit powers," *IEEE Trans. Signal Process.*, vol. 61, no. 6, pp. 1389–1403, Mar. 2013.

[22] G. Wang, H. Chen, Y. Li, and M. Jin, "On received-signal-strength based localization with unknown transmit power and path loss exponent," *IEEE Wireless Commun. Lett.*, vol. 1, no. 5, pp. 536–539, Oct. 2012.

[23] G. Wang and K. Yang, "A new approach to sensor node localization using RSS measurements in wireless sensor networks," *IEEE Trans. Wireless Commun.*, vol. 10, no. 5, pp. 1389–1395, May 2011.

[24] Y.-Y. Cheng and Y.-Y. Lin, "A new received signal strength based location estimation scheme for wireless sensor network," *IEEE Trans. Consum. Electron.*, vol. 55, no. 3, pp. 1295–1299, Aug. 2009.

[25] S. Srinivasa and M. Haenggi, "Path loss exponent estimation in large wireless networks," in *Proc. Inf. Theory Appl. Workshop*, 2009, pp. 124–129.

[26] G. Mao, B. D. O. Anderson, and B. Fidan, "Path loss exponent estimation for wireless sensor network localization," *Comput. Netw.*, vol. 51, no. 10, pp. 2467–2483, Jul. 2007.

[27] Y. Hu and G. Leus, "Self-estimation of path-loss exponent in wireless networks and applications," *IEEE Trans. Veh. Technol.*, vol. 64, no. 11, pp. 5091–5102, Nov. 2014.

[28] Y. Hu and G. Leus, "Directional maximum likelihood self-estimation of the path-loss exponent," in *Proc. IEEE Int. Conf. Acoust., Speech, Signal Process.*, Mar. 2016, pp. 3806–3810.

[29] N. Salman, M. Ghogho, and A. Kemp, "On the joint estimation of the RSS-based location and path-loss exponent," *IEEE Wireless Commun. Lett.*, vol. 1, no. 1, pp. 34–37, Feb. 2012.

[30] S. Tomic, M. Beko, and R. Dinis, "RSS-based localization in wireless sensor networks using convex relaxation: Noncooperative and cooperative schemes," *IEEE Trans. Veh. Technol.*, vol. 64, no. 5, pp. 2037–2050, May 2015.

[31] M. Gholami, R. Vaghefi, and E. Strom, "RSS-based sensor localization in the presence of unknown channel parameters," *IEEE Trans. Signal Process.*, vol. 61, no. 15, pp. 3752–3759, Aug. 2013.

[32] C. Liang and F. Wen, "Received signal strength-based robust cooperative localization with dynamic path loss model," *IEEE Sensors J.*, vol. 16, no. 5, pp. 1265–1270, Mar. 2016.

[33] B.-C. Liu and K.-H. Lin, "Distance difference error correction by least square for stationary signal-strength-difference-based hyperbolic location in cellular communications," *IEEE Trans. Veh. Technol.*, vol. 57, no. 1, pp. 227–238, Jan. 2008.

[34] B.-C. Liu, K.-H. Lin, and J.-C. Wu, "Analysis of hyperbolic and circular positioning algorithms using stationary signal-strength-difference measurements in wireless communications," *IEEE Trans. Veh. Technol.*, vol. 55, no. 2, pp. 499–509, Mar. 2006.

[35] Y. Chan and K. Ho, "A simple and efficient estimator for hyperbolic location," *IEEE Trans. Signal Process.*, vol. 42, no. 8, pp. 1905–1915, Aug. 1994.

[36] K. Ho, "Bias reduction for an explicit solution of source localization using TDOA," *IEEE Trans. Signal Process.*, vol. 60, no. 5, pp. 2101–2114, May 2012.

[37] K. Cheung, H. So, W.-K. Ma, and Y. Chan, "Least squares algorithms for time-of-arrival-based mobile location," *IEEE Trans. Signal Process.*, vol. 52, no. 4, pp. 1121–1130, Apr. 2004.

[38] Y. Wang and G. Leus, "Reference-free time-based localization for an asynchronous target," *EURASIP J. Adv. Signal Process.*, vol. 2012, no. 1, pp. 1–21, 2012.

[39] T. Pong and H. Wolkowicz, "The generalized trust region subproblem," *Comput. Optim. Appl.*, vol. 58, no. 2, pp. 273–322, 2014.

[40] J. J. Mor, "Generalizations of the trust region problem," *Optim. Methods Softw.*, vol. 2, pp. 189–209, 1993.

[41] J. Mor and D. Sorensen, "Computing a trust region step," *SIAM J. Sci. Stat. Comput.*, vol. 4, no. 3, pp. 553–572, 1983.

[42] A. Beck, P. Stoica, and J. Li, "Exact and approximate solutions of source localization problems," *IEEE Trans. Signal Process.*, vol. 56, no. 5, pp. 1770–1778, May 2008.

[43] L. Vandenberghe and S. Boyd, "Semidefinite programming," *SIAM Rev.*, vol. 38, pp. 49–95, 1994.

[44] S. P. Boyd, L. El Ghaoui, E. Feron, and V. Balakrishnan, *Linear Matrix Inequalities in System and Control Theory*, vol. 15. Philadelphia, PA, USA: SIAM, 1994.

[45] M. Grant and S. Boyd, "CVX: MATLAB software for disciplined convex programming, version 2.1," Mar. 2014. [Online]. Available: http://cvxr.com/cvx

[46] M. Grant and S. Boyd, "Graph implementations for nonsmooth convex programs," in *Recent Advances in Learning and Control (Lecture Notes in Control and Information Sciences)*, V. Blondel, S. Boyd, and H. Kimura, Eds. Berlin, Germany: Springer-Verlag, 2008, pp. 95–110. [Online]. Available: http://stanford.edu/ boyd/graph_dcp.html

[47] I. Markovsky and S. Van Huffel, "Overview of total least-squares methods," *Signal Process.*, vol. 87, no. 10, pp. 2283–2302, Oct. 2007.

[48] C. Eckart and G. Young, " The approximation of one matrix by another of lower rank," *Psychometrika*, vol. 1, no. 3, pp. 211–218, 1936.

[49] G. H. Golub and C. F. Van Loan, *Matrix Computations*, 3rd ed. Baltimore, MD, USA: The Johns Hopkins Univ. Press, 1996.

[50] A. H. Sayed, A. Tarighat, and N. Khajehnouri, "Network-based wireless location: Challenges faced in developing techniques for accurate wireless location information," *IEEE Signal Process. Mag.*, vol. 22, no. 4, pp. 24–40, Jul. 2005.

[51] F. Santucci and N. Benvenuto, "A least squares path loss estimation approach to handover algorithms," in *Proc. Int. Conf. Commun. Conf. Record, Converging Technol. Tomorrow's Appl.*, vol. 2, 1996, pp. 802–806.

[52] C. C. Pu, S. Y. Lim, and P. C. Ooi, "Measurement arrangement for the estimation of path loss exponent in wireless sensor network," in *Proc. 7th Int. Conf. Comput. Convergence Technol.*, Dec. 2012, pp. 807–812.

[53] L. Razoumov and L. Greenstein, "Path loss estimation algorithms and results for RF sensor networks," in *Proc. IEEE 60th Veh. Technol. Conf.*, vol. 7, 2004, pp. 4593–4596.

[54] G. Mao, B. D. O. Anderson, and B. Fidan, "WSN06–4: Online calibration of path loss exponent in wireless sensor networks," in *Proc. IEEE Globecom*, Nov. 2006, pp. 1–6.

[55] P. Tseng, "Convergence of a block coordinate descent method for non-differentiable minimization," *J. Optim. Theory Appl.*, vol. 109, no. 3, pp. 475–494, 2001.

[56] S. M. Kay, *Fundamentals of Statistical Signal Processing: Estimation Theory*. Upper Saddle River, NJ, USA: Prentice-Hall, 1993.

**Yongchang Hu** (M'16), photograph and biography not available at the time of publication.

**Geert Leus** (F'12), photograph and biography not available at the time of publication.